%% file: cluster_ms.tex
\documentclass[twoside,english,reprint, aps, prx, superscriptaddress, amsfonts, amssymb, amsmath, a4paper]{revtex4-1}
\usepackage{lmodern}

\usepackage[T1]{fontenc}
\usepackage{color}
\usepackage[english]{babel}
\usepackage{amsmath}
\usepackage{amssymb}
\usepackage{graphicx}

\usepackage{times}
\usepackage{caption}
\usepackage{ulem}

\usepackage[unicode=true,
 bookmarks=false,
 breaklinks=true,pdfborder={0 0 0},pdfborderstyle={},backref=false,colorlinks=true]
 {hyperref}
\hypersetup{pdftitle={A deterministic source of indistinguishable photons in a cluster state},
 pdfauthor={Dan Cogan}}



\begin{document}

\title{A deterministic source of indistinguishable photons in a cluster state}
\author{Dan Cogan}
\affiliation{The Physics Department and the Solid State Institute, Technion\textendash Israel
Institute of Technology, 3200003 Haifa, Israel}
\author{Zu-En Su}
\affiliation{The Physics Department and the Solid State Institute, Technion\textendash Israel
Institute of Technology, 3200003 Haifa, Israel}
\author{Oded Kenneth}
\affiliation{The Physics Department and the Solid State Institute, Technion\textendash Israel
Institute of Technology, 3200003 Haifa, Israel}
\author{David Gershoni}
\email{dg@physics.technion.ac.il}

\affiliation{The Physics Department and the Solid State Institute, Technion\textendash Israel
Institute of Technology, 3200003 Haifa, Israel}

\begin{abstract}
 Measurement-based quantum communication relies on the availability 
of highly entangled multi-photon cluster states. The inbuilt redundancy 
in the cluster allows communication between remote nodes using repeated 
local measurements, compensating for photon losses and probabilistic 
Bell-measurements. For feasible applications, the cluster generation 
should be fast, deterministic, and its photons - indistinguishable. 
We present a novel source based on a semiconductor quantum-dot device. 
The dot confines a heavy-hole, precessing in a finely tuned external weak 
magnetic field while periodically excited by a sequence of optical pulses. 
Consequently, the dot emits indistinguishable polarization-entangled photons, 
where the field strength optimizes the entanglement. We demonstrate Gigahertz 
rate deterministic generation of $>$90\% indistinguishable photons in a cluster 
state with more than 10 photons characteristic entanglement-length.
\end{abstract}
\maketitle
\global\long\def\ket#1{\left|#1\right\rangle }%
\global\long\def\im{\operatorname{Im}}%
\global\long\def\bra#1{\left\langle #1\right|}%

A strong barrier to developing a full-fledged quantum network is the 
unavoidable photonic losses  that exponentially limit 
the success probability of any quantum communication protocol with 
distance. Overcoming such exponential scaling is possible using the
concept of quantum repeaters \cite{Briegel1998}, in which entanglement 
swapping and purification performed at intermediate nodes enable entanglement 
distribution to distant nodes with photons \cite{Duan2001,Wehner2018}.

Inspired by measurement-based quantum computation \cite{Raussendorf2001}, 
Zwerger and coworkers introduced the idea of measurement-based quantum 
repeaters \cite{Zwerger2012}. 
This network architecture uses 
multi-photon entangled states, the cluster or graph states \cite{Hein2004, Buterakos2017}, 
which can be distributed and mutually connected through 
Bell measurements. Recently, Azuma and coworkers \cite{Azuma2015} 
extended this idea by making the repeater graph exclusively photonic, 
thereby removing the need for a long-lived quantum memory. Graph states 
provide redundancy against photon loss and the probabilistic nature of 
photonic Bell measurements: If the first measurement fails, more trials will 
significantly increase the probability of success. An essential requirement 
from such graph-states is that the photons be indistinguishable. Developing 
devices capable of deterministically producing indistinguishable photonic graph 
states is, therefore, a scientific and technological challenge of the utmost importance \cite{Munro2012,Larsen2019,Li2020,Istrati2020,Besse2020}.

Single and entangled photons can be generated via the spontaneous 
emission of an optical transition of an atom or an atom-like quantum 
emitter 
\cite{Aspect1982, Akopian2006}
Artificial atoms such 
as semiconductor quantum dots (QDs) have demonstrated tremendous 
performance as they can be incorporated into electro-optical devices, 
which dovetail with the contemporary semiconductor-based industry. 
Semiconductor QD-based devices have shown unparalleled efficiency 
and rates (for a review see Ref. \cite{Senellart2017})
, placing them as the forerunner of all physical systems considered for generating 
quantum light \cite{Zwerger2012,Munro2012,Azuma2015,Buterakos2017,Appel2021}.

Of particular relevance to this work is the proposal by Lindner and Rudolph \cite{Lindner2009} 
for generating one-dimensional photonic cluster states using semiconductor QDs. 
Their scheme uses a single QD confined electron spin precessing in a magnetic field while driven 
by a temporal sequence of picosecond laser pulses. Upon excitation of the 
QD, a single photon is deterministically emitted, and its polarization 
is entangled with the state of the QD confined spin. 
This process repeats many times to generate a large 1D cluster of 
entangled photons. 

Schwartz and coworkers demonstrated a modification of this proposal by 
generating a 1D cluster state based on QD device and characterizing the cluster entanglement 
length \cite{Schwartz2016}. 
The entanglement robustness of such a photonic cluster is mainly determined 
by the ratio between the optical transition radiative rate and the spin-precession 
rate \cite{Schwartz2016} and by the coherence properties of the QD spin \cite{Cogan2018,Bechtold2015}. 
The indistinguishability between the emitted photons is mainly determined by 
the nature of the optical transition, which results in the photon emission \cite{Kiraz2004}.

In \cite{Schwartz2016}, the entangler was the spin of the QD confined 
dark exciton (DE) \cite{Schwartz2015}.
The short-range electron-hole exchange interaction removes the degeneracy between the two 
DE spin states. Therefore, a coherent superposition of the DE eigenstates 
naturally precesses, even without an external magnetic field. Unfortunately, 
this precession is not easily controlled by external means, limiting the entanglement length. 
Moreover, the emitted photon leaves the DE in an excited state, having a shorter 
lifetime than the radiative time. As a result, the emitted photons have limited indistinguishability 
\cite{Kiraz2004},
required for efficient Bell state measurements.

Here we use the heavy-hole (HH) spin \cite{Gerardot2008,Brunner2009,Cogan2018} 
as an entangler instead of the DE. The HH has a half-integer spin, and in the 
absence of a magnetic field, its two spin states are Kramers-degenerate. Therefore 
its precession rate can be finely controlled by the strength of the external field. 
As we show below, this allows the optimization of the robustness of the entanglement 
in the cluster state and determines the photon generation rate. Moreover, in the HH 
case, the photon leaves the HH in a stable ground state, and therefore, the photons 
are highly indistinguishable. \\

\begin{figure*}
\begin{centering}
\includegraphics[width=1\textwidth]{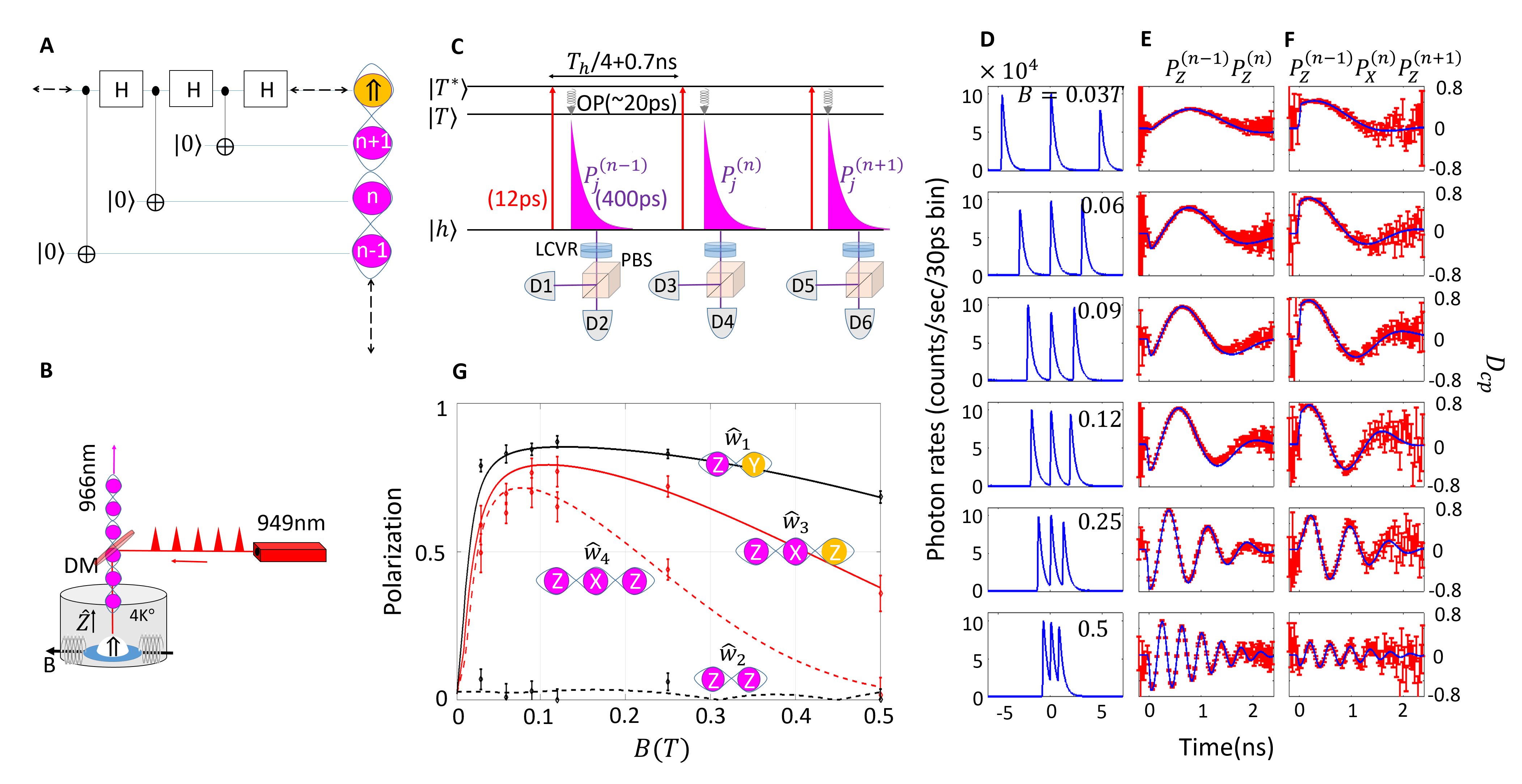}
\par\end{centering}
\caption{\label{fig: Cluster generator} {\bf The cluster witnesses.} 
({\bf A}) The cluster state is ideally generated by sequential Hadamard-and CNOT-gates. 
({\bf B}) The QD-based device. DM is dichroic-mirror. $\vec{B}=B\hat{x}$ is an external magnetic field. 
({\bf C}) The QD’s spin-configurations and transition times during the measurement. 
$\ket{h}$, $\ket{T}$, and $\ket{ T^*}$ are the HH, trion, and excited trion states. 
Red arrows are laser pulses, curly arrows are optical-phonons, and the pink 
exponential-decays are the emitted single-photons. $T_h$ is the HH spin-precession period. 
We detect two and three consequent photons and project their polarization using 
liquid-crystal-variable-retarders (LCVR) and polarized-beam-splitters (PBS). $P_{j}^{(i)}$ 
represents the polarization of the i’s-photon in the cluster, projected on the j polarization state. 
({\bf D}) Time-resolved PL measurements for various magnetic fields.
({\bf E}) [({\bf F})] 2-photon (3-photon) correlations. Red marks are the time-resolved 
measurement of the last correlated photon’s degree-of-circular-polarization ($D_{cp}$) 
as a function of time from the second (third) pulse. Blue lines represent the best fitted 
central-spin-evolution model for the positive-trion \cite{Cogan2020,Cogan2018}. 
({\bf G}) The measured (error bars) and calculated (color matched lines) cluster-state 
witnesses as deduced from the fits in (E) and (F) vs. the in-plane magnetic field strength. 
Strings of polarized photons (pink-) and spin (yellow-) circles describe the witnesses.}
\end{figure*}

\begin{figure*}
\begin{centering}
\includegraphics[width=1\textwidth]{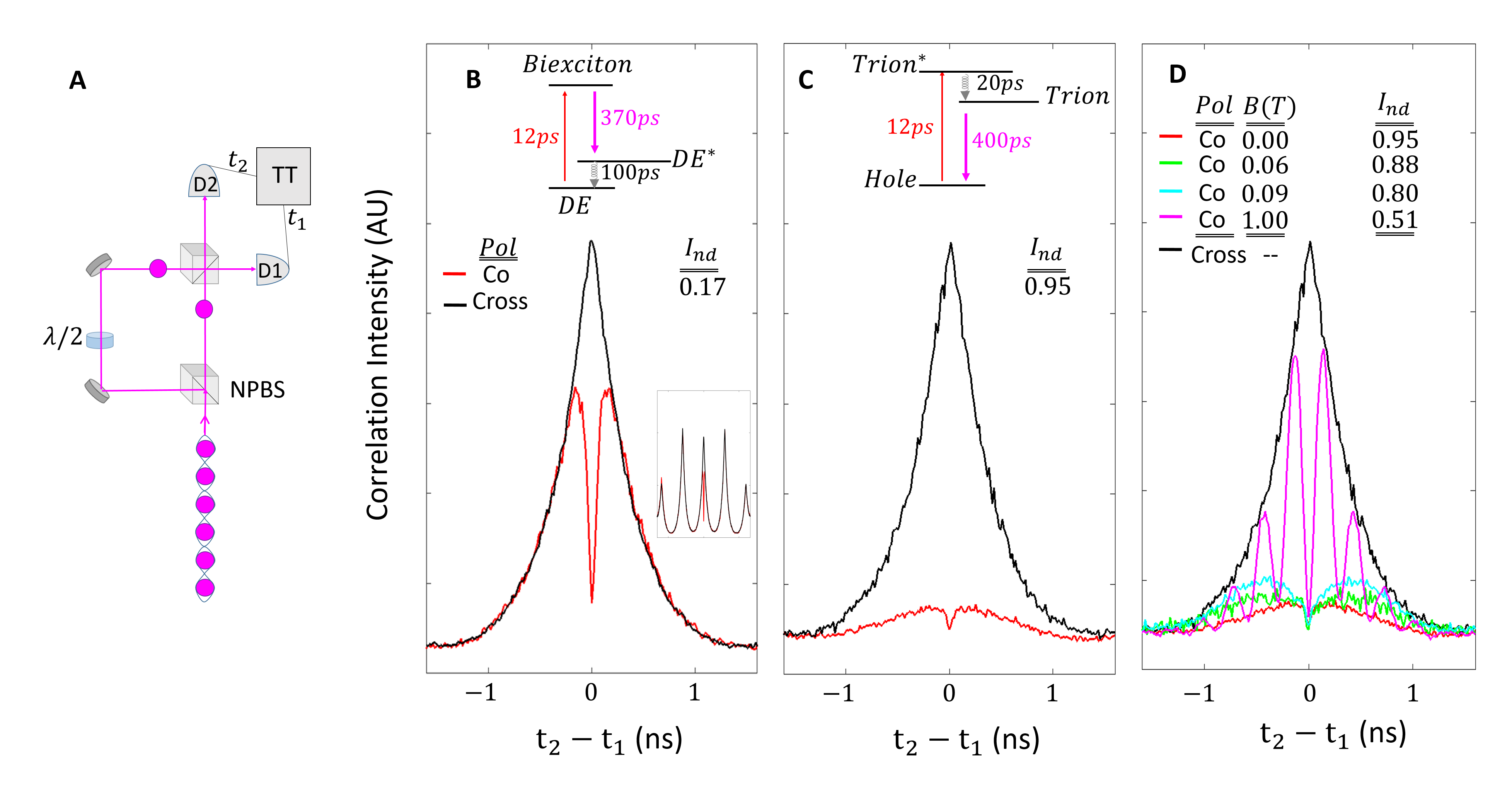}
\par\end{centering}
\caption{\label{fig:Indistinguishability}{\bf Photon indistinguishability.} 
{({\bf A})} The interferometer setup. Two consequent cluster-photons interfere on 
a Non-Polarizing-Beam-Splitter (NPBS). Detectors on the two output ports register the 
detection times $t_{1}$($t_{2})$ of the photons. A half-wave plate ($\lambda/2$) 
in a rotation mount selects between the two photons' co- or cross-linear polarization state. 
({\bf B}) {[}({\bf C}){]} Indistinguishability measurements between two consecutive 
photons from a cluster where the DE [HH] is the entangler. The solid lines represent 
the measured second-order correlation function $g^{(2)}(t_{2}-t_{1})$. The measured 
indistinguishability is defined as $I_{nd}=1-A_{co}/A_{cross}$, where $A_{co}$($A_{cross}$) 
is the area under the co-(cross-) polarized correlation function, represented by the red (black) 
solid line. The inset in (B) [(C)] describes the energy levels and the transition rates between the 
levels for the DE [HH]. The bottom inset in (B) describes $g^{(2)}(t_{2}-t_{1})$ on a 13ns time scale. 
The interference is visible on the zero-time difference coincidences. 
({\bf D}) Indistinguishability measurements, like in (C) for various externally applied magnetic fields.}
\end{figure*}

We follow Ref. \cite{Lindner2009}, as schematically described in Fig.~1A, by applying 
a periodic sequence of CNOT- and Hadamard- gates on the HH for producing  the cluster state.
Fig.~1B and Fig.~1C schematically describe our QD-based device operation and 
the HH-positive trion energy levels. Each laser pulse (red upward arrow) excites 
the confined HH ($\ket{\Uparrow}$) to the excited- positive trion state ($\ket{\Uparrow\Downarrow\uparrow}${*}), 
in which the electron resides in its respective second energy level. The electron 
decays to the trion ground level within about 20~ps by emitting a spin-preserving 
optical phonon \cite{Benny2012}. The trion then recombines within 
about 400~ps, by emitting a photon (marked in pink), leaving the HH at its ground 
level. A dichroic mirror steers the emitted photons to the detectors.

Both the HH and the positive trion act as spin qubits \cite{Loss1998}.
The selection rules for the optical transitions 
associated with the excitation and emission result in entanglement between the 
photon polarization and the HH-spin polarization \cite{SM}. Therefore each excitation-emission 
step is an actual realization of a CNOT gate between the spin and the photon qubits.
We realize the Hadamard gate on the spin qubit by timing the temporal precession 
of the HH to match one-quarter of the its precession-period plus a small 
addition compensating for the finite radiative time of the trion. A sequence of 
resonantly tuned linearly polarized laser $\pi-$pulses experimentally realizes the 
generation of the cluster state (Fig.~1C). Each pulse results in the addition of an 
entangled photon to the cluster state.

We measure the cluster state’s entanglement robustness for various external 
magnetic fields using three cycles of the cluster protocol and considering 
events in which three consequent photons are detected. In Fig.~1D, 
we present these measurements for six different magnetic field strengths. We 
use the last detected photon for the tomography of the HH-spin. The tomography 
utilizes time-resolved degree-of-circular-polarization ($D_{cp}$) measurement for 
monitoring the trion's spin evolution during its radiative decay
back to the HH \cite{Cogan2020,Cogan2021}.

Fig.~1E and Fig.~1F show the $D_{cp}(t)$ of the correlated photon-spin states 
and photon-photon-spin states, respectively, measured for six different magnetic fields. 
From the fit to these measurements, we extract the following four cluster-state witnesses: 
$\hat{w}_{1}=P_{Z}^{(n-1)}S_{Y}$, $\hat{w}_{2}=P_{Z}^{(n-1)}P_{Z}^{(n)}$,
$\hat{w}_{3}=P_{Z}^{(n-1)}P_{X}^{(n)}S_{Z}$ and $\hat{w}_{4}=P_{Z}^{(n-1)}P_{X}^{(n)}P_{Z}^{(n+1)}$,
where $P_{j}^{(i)}$ represents the polarization projection of the i’s-photon in the string, 
on the j polarization, and $S_{j}$ is the polarization of the HH-spin, projected on the j state. 
We note that $\hat{w}_{1}$ and $\hat{w}_{3}$ are directly extracted from the 
polarization of correlated detection of 2- and 3-photons, shown in Fig.~1E and 1F,
respectively, while $\hat{w}_{2}$ and $\hat{w}_{4}$ are extracted from the temporal 
integration of these measurements. For an ideal cluster-state (containing ideal CNOT and Hadamard gates), 
$\hat{w}_{1},\hat{w}_{3}$, and $\hat{w}_{4}$ equal 1, and $\hat{w}_{2}$ equals 0.

In Fig.~1G, we present the four measured witnesses as a function of the magnetic field strength. 
The solid lines are the calculated witnesses using our state-evolution-model \cite{SM}. 
The parameters used for the model calculations were independently measured and listed in 
Ref. \cite{Cogan2021}. Very good quantitative agreement between 
the measured data and the calculations is obtained.\\ 

We now turn to measure the indistinguishability between two sequential photons in 
the cluster state. We use the HOM-setup \cite{Hong1987,Santori2002} illustrated in 
Fig.~2A to compare the industinguishability resulting from using the DE as entangler (Fig.~2B) with that of the HH (Fig.~2C). 
The indistinguishability is given by the ratio between the second-order interference 
of co- to cross-polarized photons. At zero magnetic field, the DE-cluster photons 
indistinguishability amounts to 17\%, much lower than that of the HH-cluster, which amounts to 95\%.

The indistinguishability between photons emitted from a quantum source of single-photons 
depends on the initial and final states\textquoteright{} temporal stability \cite{Kiraz2004}.
The insets to Fig.~2B and 2C illustrate the energy level structures and the relevant 
transition rates for the DE and the HH, respectively. This  
significant improvement in the indistinguishability of the HH-cluster photons results 
from the temporal stability of the HH final ground level compared with the excited level of the DE \cite{SM}.

The externally applied magnetic field reduces the indistinguishability of the HH-cluster photons.
Fig.~2D presents indistinguishability measurements of the HH-cluster photons for 
various magnetic fields. As the field increases, the indistinguishability decreases due 
to the field-induced spectral broadening. At large fields, when the Zeeman splitting 
surpasses the radiative linewidth \cite{SM}, the indistinguishability drops to 50\%, and quantum 
beats are observed in the time-resolved correlation measurements \cite{Legero2004}.\\

\begin{figure}
\begin{centering}
\includegraphics[width=1\columnwidth]{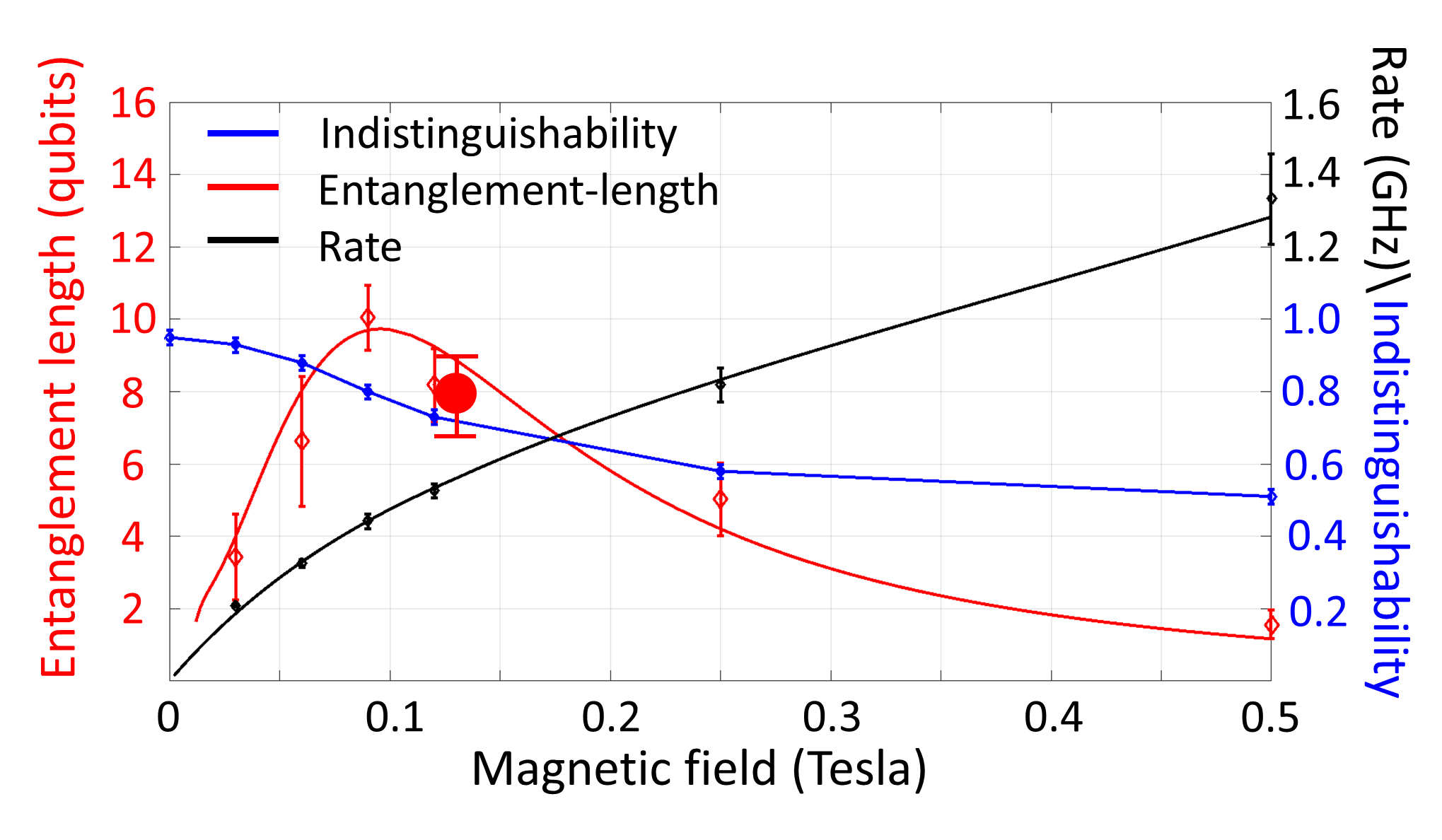}
\par\end{centering}
\caption{\label{fig:cluster_parameters} {\bf Cluster characterization.} 
The cluster state’s localizable-entanglement (LE)
characteristic length (red), photon-indistinguishability (blue), and 
generation rate (black) as a function of the magnetic field. Marked 
points and error bars are deduced from our measurements. The lines 
represent calculations in which the process map was obtained from our 
state-evolution model \cite{SM}. The red circle stands for the LE length 
at B=0.12T, where the process map was measured.}
\end{figure}

Fig.~3 summarizes the indistinguishability properties of the cluster photons, the 
robustness of the generated entanglement, and the photon-generation-rate as 
a function of the applied magnetic field. 

The entanglement robustness is quantified using localizable entanglement (LE) 
\cite{Verstraete2004}, defined as the magnitude of the entanglement between 
two qubits in the cluster after all the other qubits are projected on suitable 
bases. The LE decays exponentially with the distance between the two qubits \cite{Verstraete2004,Schwartz2016}. 
Thus $\zeta_{LE}$, defined as the characteristic decay-length of the LE, is a figure-of-merit characterizing 
the robustness of the entanglement in the cluster.

Fig.~3 presents $\zeta_{LE}$ vs. the magnetic field strength, where $\zeta_{LE}$ is 
deduced from the ratio between two of the measured witnesses (red diamond marks) according to:
\begin{equation}
\zeta_{LE}=-1/\ln\left(\hat{w}_{3}/\hat{w}_{1}\right).\label{eq:entanglement-length-1}
\end{equation}
The difference between these two witnesses is simply the $D_{cp}$ loss due to one 
application of the process map. For a field strength of $0.12$ Tesla, 
$\zeta_{LE}$ (full red circle) was deduced from a full measurement of the process map \cite{SM}. 
the measured $\zeta_{LE}$  vs. magnetic field is compared with our 
central-spin-evolution model (solid red line), using the measured QD-parameters \cite{Cogan2021}.

Fig.~3 clearly shows an optimum for the entanglement robustness at Bx=0.09 Tesla, 
in which $\zeta_{LE}=10$. For stronger fields, the precession period becomes shorter. 
As a result, the ratio between the radiative-time and precession-time becomes larger, 
thereby increasing the deviation of the 2-qubit gate from the ideal CNOT gate. 
For weaker fields, the precession-time increases. Consequently, the 
time-difference between consecutive pulses increases to a level in which the  HH-spin dephasing becomes significant.  

To complete the device characterization,  Fig.~3 presents also the measured 
indistinguishability (in blue) and the cluster generation rate (in black) vs. the magnetic field strength.
The indistinguishability monotonically decreases from 95\% as the field increases. 
This decrease is due to the field-induced broadening of the optical transition. At 
$B=0.09$ Tesla, the measured indistinguishability is better than 80\%. 

The time between consecutive laser pulses is set to approximately match a quarter 
of a precession period. Therefore, the generation rate monotonically increases as 
a function of the magnetic field. At the optimal field of 0.09 Tesla, the generation rate is about 0.5 GHz.

The overall detection efficiency of our system is currently better than 1\%. 
The main losses are due to limited light-harvesting efficiency from the planar 
microcavity containing the QD (\textasciitilde 20\%), free space into fiber coupling 
(\textasciitilde 50\%), detectors quantum efficiency (\textasciitilde 80\%) and the 
overall optical-elements transmission efficiency (\textasciitilde 10\%). With the 
current system efficiency, we measure four-photon correlations at a few Hertz rate. 
This detection rate is enough to demonstrate and characterize our cluster-state-generating-device. 

There is an obvious need to improve the collection efficiency. A clear way is to 
integrate the QD into a 3D photonic microcavity \cite{Senellart2017,Liu2018,Tomm2021}. 
Using such a microcavity with a better mode matching to a single-mode fiber can 
increase the system efficiency by at least an order of magnitude. In addition, 
a 3D microcavity can significantly shorten the radiative lifetime due to the Purcell effect. This will significantly improve both 
the entanglement length and the photon indistinguishability.\\

In summary, we demonstrate a-Gigahertz rate deterministic-generation of a cluster 
state of indistinguishable photons using a QD confined HH-spin as an entangler. 
We show that by optimizing the in-plane external magnetic field, we achieve a 
characteristic entanglement-length of 10 photons, and photon indistinguishability 
which is better than 80\%. Further feasible optimizations of the device can bring 
widespread measurement-based implementations of quantum communication and information processing closer.

\section*{Acknowledgments}
The support of the Israeli Science Foundation (ISF), and that of the European 
Research Council (ERC) under the European Union\textquoteright s Horizon 2020 research and 
innovation programme (Grant Agreement No. 695188) are gratefully acknowledged.

\nocite{Cogan2020}

\input{cluster_ms.bbl}

\bibliographystyle{aipnum4-1}

\end{document}


\title{Supplemental material for "A deterministic source of indistinguishable
photons in a cluster state''}
\author{Dan Cogan}
\affiliation{The Physics Department and the Solid State Institute, Technion\textendash Israel
Institute of Technology, 3200003 Haifa, Israel}
\author{Zu-En Su}
\affiliation{The Physics Department and the Solid State Institute, Technion\textendash Israel
Institute of Technology, 3200003 Haifa, Israel}
\author{Oded Kenneth}
\affiliation{The Physics Department and the Solid State Institute, Technion\textendash Israel
Institute of Technology, 3200003 Haifa, Israel}
\author{David Gershoni}
\email{dg@physics.technion.ac.il}

\affiliation{The Physics Department and the Solid State Institute, Technion\textendash Israel
Institute of Technology, 3200003 Haifa, Israel}

\maketitle
\tableofcontents

\section*{SM1. The sample and optical system}

In the heart of our device is an InGaAs self assembled semiconductor quantum dot (QD), 
as illustrated in Fig. 1B in the main article. The QD contains a confined heavy-hole (HH) 
spin qubit \cite{Loss1998,DiVincenzo_2000,Gerardot2008,Brunner2009,Cogan2018}. 
We define the QD's shortest dimension (\texttildelow 3nm), the growth direction, and 
the spin quantization axis as the Z-axis. The externally applied magnetic field direction is 
defined as the X-axis. The QD is embedded in a planar microcavity, formed by 2 
Bragg-reflecting mirrors, facilitating light-harvesting efficiency of about 20\% by a 0.85 numerical aperture objective above the QD.

The experimental system  is described in Ref.  \cite{Cogan2018}. A modified version is schematically described in Fig.~S1. 
The pulsed laser light  is resonantly tuned to the HH-excited trion optical transition \cite{Cogan2021}, 
and its power is set to a $\pi-$area intensity. The polarization of the light is  
computer-controlled using pairs of liquid crystal variable retarders (LCVRs) and polarizing  beam splitters (PBS). 
Very weak above bandgap CW light is also used to stabilize the charge state of the QD \cite{Cogan2018}.  
Similar sets of LCVRs and PBSs are used to project the polarization of the collected photons 
on 6 different polarization bases. We use highly efficient 
transmission gratings to spectrally filter the emitted photons. The filtered photons are detected 
using efficient (>80\%) single-photon superconducting detectors, with temporal resolution 
of about 30ps. The photons detection times are recorded using time tagging correlation module.

\begin{figure*}
\begin{centering}
\includegraphics[width=0.7\textwidth]{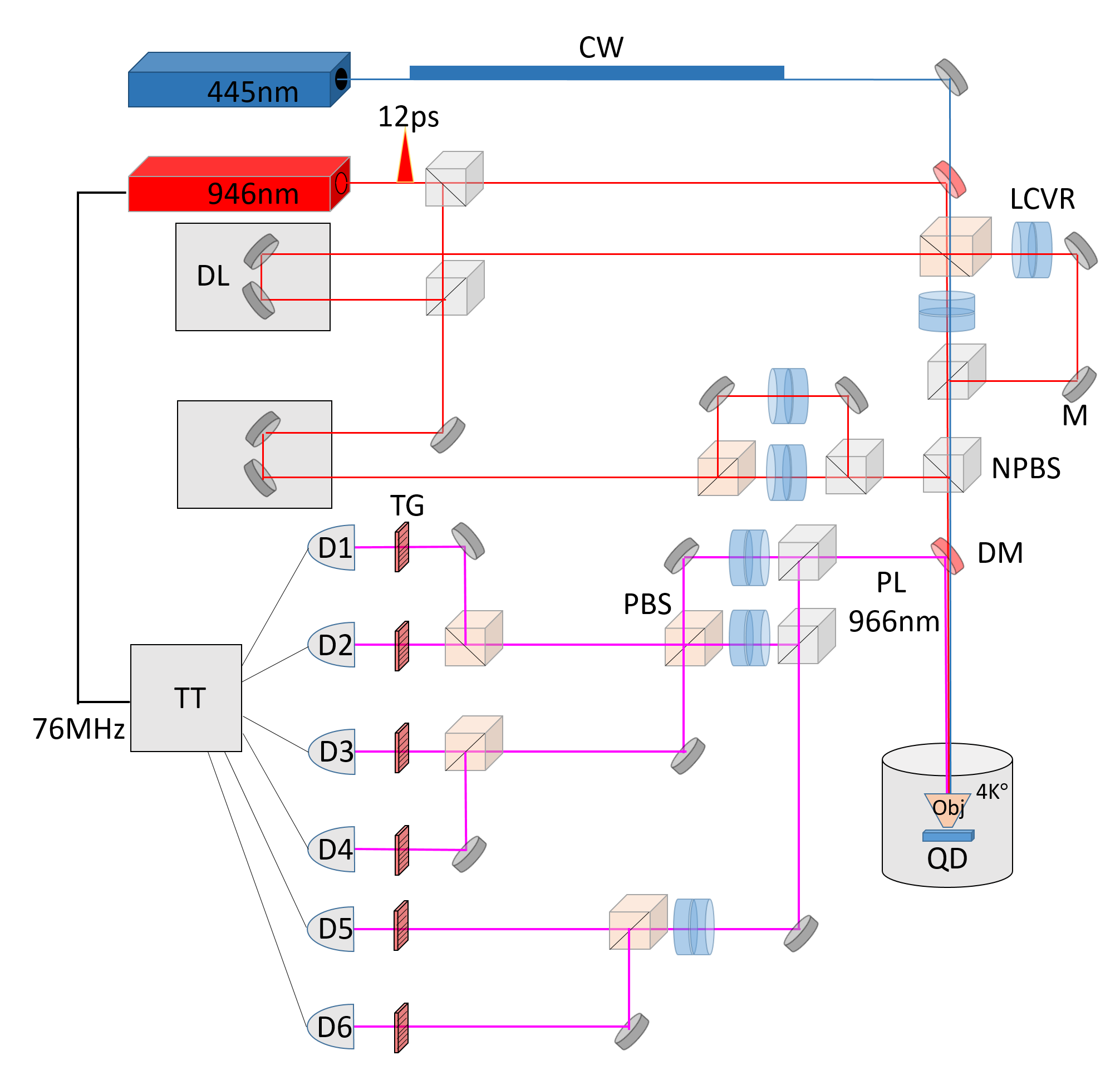}
\par\end{centering}
\caption{{\bf The experimental system.} CW, continuous wave; DL, Delay line; M, mirror; DM, dichroic mirror; 
NPBS, non-polarizing beam splitter; PBS, polarizing beam splitter; OBJ, microscope objective; LCVR, liquid crystal 
variable retarder; Dn, single-photon detector n; TT, Time tagging correlation module;}
\end{figure*}

\section*{SM2. The state-evolution model}

Here we outline the state evolution model, which describes the actual single cycle of the periodic protocol. 
The cycle is ideally composed of two gates. A CNOT gate between the QD spin-qubit and the emitted photon 
polarization-qubit, followed by a Hadamard gate acting on the spin qubit only.
The CNOT gate is realized by the pulsed optical excitation of the HH spin and the resulting single-photon 
emission. The Hadamard gate is realized by the timed free coherent precession of the HH around the direction of the externally 
applied magnetic field. We developed the state evolution model in order to realistically describe the process map and its deviation 
from the ideal map, composed of the ideal gates.

The optical excitation of the HH results in a positively charged trion composed of two paired holes and a single electron 
(Fig.~S2A). The two holes and the single electron spins interact via the hyperfine interaction with nuclear spins in their 
vicinity and with the externally applied magnetic field. In Ref. \cite{Cogan2021}, we developed a model that quantitatively 
describes both spins' evolution under these conditions, focusing particularly on the relevant regime, where the external field 
is comparable in magnitude to one standard deviation of the effective nuclear magnetic field (Overhauser field) due to random 
statistical fluctuations in the nuclei spins \cite{Efros2002}.

In Ref.  \cite{Cogan2021}, we show that under these conditions the central spin $\vec{S}(t)$ evolution is given by: 

\begin{align} 
\vec{S}(t)=\bar{G}\vec{S}_{0}\label{eq:1}
\end{align}
where $\vec{S}_{0}=\vec{S}(0)$ is the central spin initial value
and $\bar{G}$ is a 3 \texttimes{} 3 tensor whose elements can be calculated numerically as explained  in Ref \cite{Cogan2021}.
We mark by $\bar{G}^{HH}$ and $\bar{G}^{trion}$ the tensors for the HH and trion.

Each cycle starts with an initial HH spin state $\vec{S}_{0}$, which is the state in which the HH was left at the end of
the previous cycle. The photons emitted by the QD during previous cycles can be safely disregarded. 

A 12ps $\pi$-area laser pulse converts the HH state to an excited
positive trion state. The optical transitions between these two
spin qubits are governed by the following $\Pi$ -system optical polarization selection
rules \cite{Cogan2018}: 
\begin{align}
\ket{\Uparrow} & \stackrel{\ket{-Z}}{\longleftrightarrow}\ket{\Uparrow\Downarrow\uparrow},\nonumber \\
\ket{\Downarrow} & \stackrel{\ket{+Z}}{\longleftrightarrow}\ket{\Downarrow\Uparrow\downarrow}.\label{eq:Selection_rules-1-1-1}
\end{align}
where 
$\ket{+Z}$ and $\ket{-Z}$ are photons with right and left circular polarizations or angular momentum
of +1 and -1, respectively along the z-direction. Since our short laser pulses can be considered instantaneous it follows that a short laser
pulse polarized $\ket{+X}=\left(\ket{+Z}+\ket{-Z}\right)/\sqrt{2}$, coherently converts the HH spin state into a similar trion state\cite{Cogan2020,Cogan2021}.
described by $\vec{S}_{0}$. 
The trion then radiatively decays exponentially ( $\sim e^{-t/\tau_{photon}}$ with $\tau_{photon}=400~ps$) into an entangled HH+photon state. 

We proceed by separating the temporal evolution into three
domains: the trion evolution before the emission, the emission itself
and the HH evolution after the emission. 
The evolution during the first and third domains is described by Eq.~\ref{eq:1}.  In the following we discuss the second part which includes both qubits: the emitted photon and the spin. 


Note that while above we used $\vec{S}$ to describe the QD spin,
an equivalent description in terms of a density matrix is 
\begin{equation}
\hat{\rho}_{2\times2}=\frac{1}{2}\left(\hat{\sigma}_{0}+S_{x}\hat{\sigma}_{x}+S_{y}\hat{\sigma}_{y}+S_{z}\hat{\sigma}_{z}\right)
\end{equation}
with $\hat{\sigma}_{i}$ being the Pauli matrices and $\hat{\sigma}_{0}$ the identity matrix. 
According to the polarization selection rules for the optical transition between the
trion and the HH spins, described by Eq.~\ref{eq:Selection_rules-1-1-1},
a photon emission occurring right after the moment $t=t^{'}_{+}$, results in the spin+photon state 
\begin{equation}
\hat{\rho}_{4\times4}(t'_{+})=\left[\begin{array}{ccc}
0\\
 & \hat{\rho}_{2\times2}(t'_{-})\\
 &  & 0
\end{array}\right].
\end{equation}
Projecting this on the photon state $\frac{1}{2}(I+\vec{m}\cdot\vec{\sigma})$,
where $\vec{m}$ is the photon's normalized Stokes vector, we obtain
after tracing over the photon, the HH state 
\begin{equation}
\frac{1}{2}\left[\begin{array}{cc}
(1+S_{z})(1-m_{3}) & (S_{x}-iS_{y})(m_{1}-im_{2})\\
(S_{x}+iS_{y})(m_{1}+im_{2}) & (1-S_{z})(1+m_{3})
\end{array}\right].
\end{equation}
where $\vec{S}=\vec{S}^{trion}(t'_{-})$ right before the photon emission. The (trace) normalization
of this gives the probability $p=\frac{1}{2}(1-S_{z}m_{3})$ to find
the photon $\hat{m}$-polarized and the expectation value of $\vec{\sigma}$
gives the HH spin 
\begin{equation}
p\vec{S}^{HH}(t'_{+})=\frac{1}{2}(m_{1}S_{x}-m_{2}S_{y},m_{1}S_{y}+m_{2}S_{x},S_{z}-m_{3})\label{TM}
\end{equation}
We may write this as $p\vec{S}^{HH}(t'_{+})\equiv\bar{T}^{_{m}}\vec{S}^{trion}(t'_{-})$
where $\bar{T}^{_{m}}$ is defined by Eq.~\ref{TM}. Note that $\bar{T}^{_{m}}$
is an affine rather than a linear transformation.


Collecting everything together, this suggests that the spin at the end
of the cycle (occurring at the moment $t_{pulse}$ of the next pulse)
should satisfy 
\begin{align}
 & P^{(m)}\left\langle \overrightarrow{S}(t_{pulse})\right\rangle =\frac{1}{\tau_{photon}}(1-e^{-\frac{t_{pulse}}{\tau_{photon}}})\label{eq:S(t) result}\\
 & \int d{\cal P}\int_{0}^{t_{pulse}}e^{-\frac{t'}{\tau_{photon}}}dt'\bar{G}^{HH}(t_{pulse}-t')\bar{T}^{_{m}}\bar{G}^{trion}(t')\vec{S}_{0}\nonumber 
b\end{align}
where 
\begin{equation} d{\cal P}=\frac{1}{(2\pi)^{3/2}}\exp\left(-\frac{1}{2}B^{2}\right)d^{3}B\label{eq:dP}
\end{equation} is the probability element of the normalized nuclear field \cite{Cogan2021}, and   
$P^{(m)}$ is the probability to find the emitted photon $\hat{m}$-polarized.
$P^{(m)}$ may be obtained from 
\begin{align}
 & 1-2P^{(m)}=\frac{m_{3}}{\tau_{photon}}(1-e^{-\frac{t_{pulse}}{\tau_{photon}}})\label{P(t)}\\
 & \int d{\cal P}\int_{0}^{t_{pulse}}e^{-\frac{t'}{\tau_{photon}}}dt'\left(\hat{z}\cdot\bar{G}^{trion}(t')\vec{S}_{0}\right).\nonumber 
\end{align}
In practice $e^{-\frac{t_{pulse}}{\tau_{photon}}}\ll1$ may be safely
neglected.


We solve the integrals in Eq.~\ref{eq:S(t) result},\ref{P(t)} numerically\footnote{For certain specific photon polarizations $\hat{m}$, one may show
analytically that some components of the integral vanish by reflection
symmetries.} with the QD carriers' g-factor  and hyperfine  tensors 
as measured independently and described elsewhere 
\cite{Cogan2021}. 

We use this model to calculate the expected values of the four
cluster-state witnesses displayed in Fig.~1G of the main article. 
Similarly, we use the model for calculating the expected process map \cite{Cogan2021a} and the localizable entanglement
characteristic length \cite{Verstraete2004,Schwartz2016} as a function of the externally applied magnetic field,
as displayed in Fig.~3 of the main article. 
As can be seen in both figures the model quite accurately describes the measurements. 



\section*{SM3. Measuring the process map of the repeated cycle of the protocol}

\begin{figure*}
\begin{centering}
\includegraphics[width=0.7\textwidth]{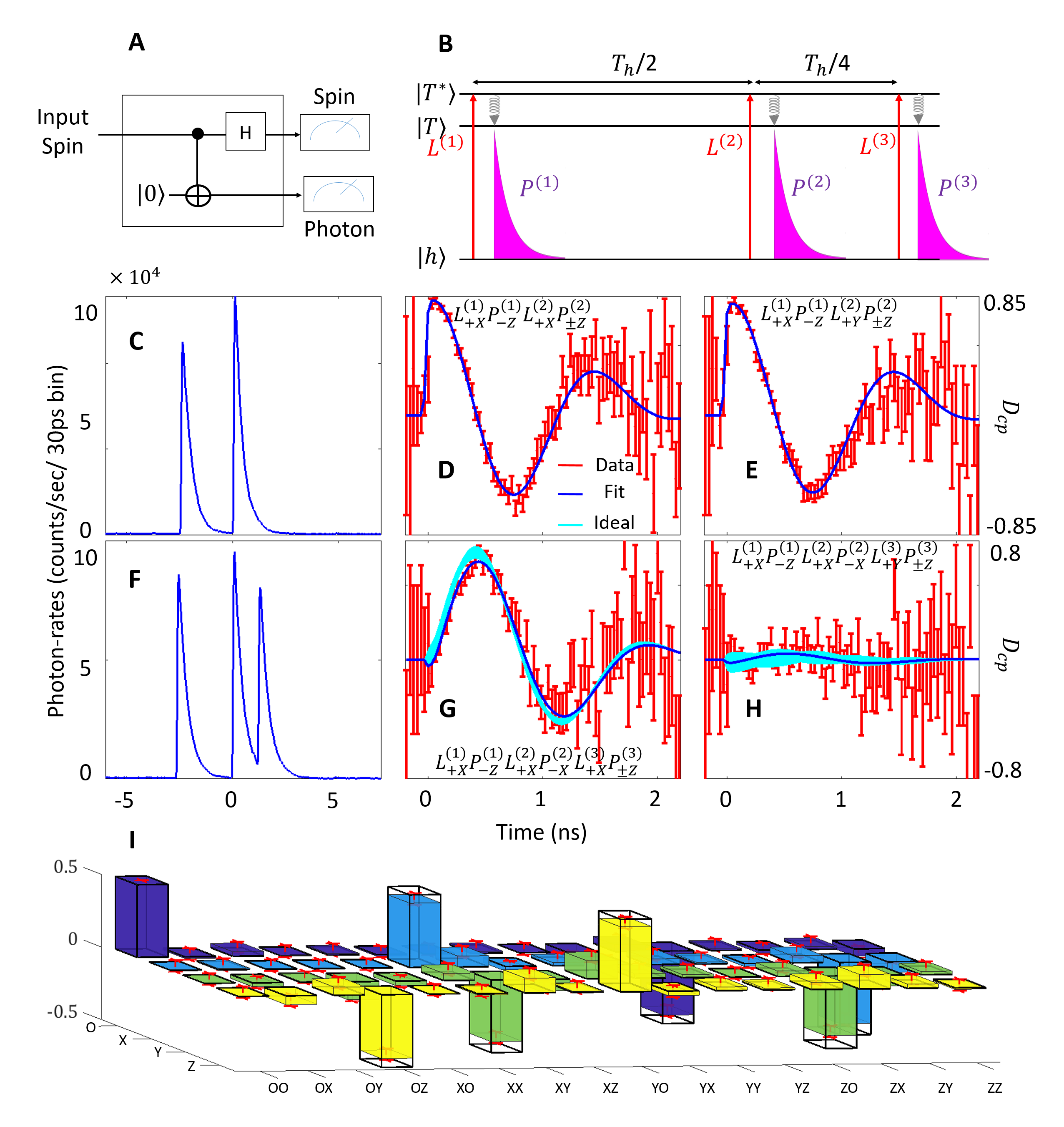}
\par\end{centering}
\caption{\label{fig:process}{\bf Measurement of the process map for an applied
magnetic field of 0.12T.} ({\bf A}) Schematics description of the process map acting on the HH spin. 
({\bf B}) Schematic description of the pulse sequence and resulted photon emission
used for measuring the process map elements for initialized HH spin $\protect\ket{\pm Z}$
states. $L_{i}^{(n)}$ ( $P_{i}^{(n)}$) mark the n's laser
pulse (single-photon detected) where i marks the polarization.  
({\bf C, D-E}) Time resolved emission and degree-of-circular-polarizations
($D_{cp}$) measurements for characterizing the HH $\protect\ket{+Z}$
input spin state. ({\bf F, G-H}) Time resolved emission and $D_{cp}$ measurements for
characterizing the spin+photon state, resulting from application
of the process map on the $\protect\ket{+Z}$ input spin state. The
red error-bars constitute one standard deviation of the experimental
uncertainty. The blue lines represent the best fitted model results \cite{Cogan2020}. 
The light-blue lines represent the expected results assuming ideal process map. ({\bf I})
The measured (filled colored bars) and ideal (empty bars) process
map. X, Y, Z, and O stands for the corresponding Pauli and identity
matrices basis representation, respectively.}
\end{figure*}

As discussed above, the cluster state is generated by repeatedly applying the same process on the QD confined spin.
 This means that one can fully characterize the cluster quantum state for any number of qubits
if the process-map is accurately known \cite{Schwartz2016,Cogan2021a}. 
The process map,  which is schematically described in Fig.~\ref{fig:process}A,
maps the spin qubit state into an entangled spin-photon two qubit state. Since the spin quantum state can be fully described by a 2x2 density matrix
and the spin+photon state by a 4x4 density matrix,  it follows that  the process map can be fully described by a 4x16
positive and trace-preserving matrix which contains 64 real matrix elements, as presented in Fig.~\ref{fig:process}A.

In Ref.  \cite{Cogan2021a} we describe a novel way for obtaining the process map from a set of time resolved 
polarization sensitive two and three photon correlation measurements. 
In Fig.~\ref{fig:process}B. we schematically describe the experimental method where we used a 2-pulse-experiment  to initialize and HH spin qubit and then to 
measure the resulting HH quantum state  \cite{Cogan2020}. Similarly, a 3-pulse-experiment is used to initialize the HH
spin, apply one cycle of the process on the initialized spin and then to measure the resulted entangled spin+photon state.

In Fig.~\ref{fig:process}C we present time-resolved two-pulse-correlation-measurements
in which the input HH was initialized to $\ket{+Z}$ state and then its state was measured \cite{Cogan2020}.
Fig.~\ref{fig:process}D and E show the measured time resolved degree of circular polarization ($D_{cp}(t)$) of the emission resulting from 
a second pulse polarized +X and  +Y respectively. 

Similarly, we initialize the HH-state to six different states from three orthogonal
bases \cite{Schwartz2015a,Cogan2021a}. For
each of those 6 states, we used 2 $D_{cp}(t)$ two-photon correlation measurements (not shown)
like the measurements displayed in Fig.~\ref{fig:process}D and E for tomography.

Fig.~\ref{fig:process}F shows time-resolved three-pulse-correlation-measurement
where one cycle of the process is applied to the $\ket{+Z}$ input
state and the resulting spin+photon state is measured. In this experiment,
the second photon is projected on the -X polarization. Fig. \ref{fig:process}G
and H show the time resolved $D_{cp}(t)$ of the three-photon correlations measurements resulting from 
a third pulse polarized +X and +Y, respectively \cite{Cogan2020}. 
 
Fig.~\ref{fig:process}I shows the results of the full-tomographic
measurements of the process map. For acquiring these measurements,
for the 6 HH spin initializations, we applied one cycle of the protocol 
and measured the resulting spin+photon state by projecting both the
spin and the photon on different orthogonal polarization bases \cite{James2001}.
For characterizing each of these 6 spin+photon states, we used 12
$D_{cp}(t)$ 3-photon correlations measurements like the ones presented
in Fig.~\ref{fig:process}G and H.

Lastly, in order to obtain the physical process map that best fits our measured results,
we use a novel minimization method \cite{Cogan2021a}. 
In Fig.~\ref{fig:process}I we compare  between the obtained process-map
and the ideal one. The fidelity \cite{Jozsa1994,Schwartz2016}
of the measured map to the ideal one is 0.9.

\section*{SM4. Estimation of the photon indistinguishability. }

The indistinguishability between photons emitted from a quantum source of single-photons 
depends on the initial and final states'\textquoteright{} temporal stability. The indistinguishability, 
can be roughly estimated by \cite{Kiraz2004,Gantz2017}:
\begin{equation}
I_{nd}=[(1+\tau_{photon}/\tau_{final})(1+\tau_{init}/\tau_{photon})]^{-1}\label{eq:Ind_model}
\end{equation}
In Eq.~\ref{eq:Ind_model}, $\tau_{init}$ is the initial state\textquoteright s generation-time (``jitter''), 
$\tau_{final}$ is the final state\textquoteright s lifetime, and $\tau_{photon}$ is the 
photon\textquoteright radiative lifetime. To achieve high indistinguishability, 
$\tau_{init}/\tau_{photon}$ should be minimized while $\tau_{final}/\tau_{photon}$ 
should be maximized.

The insets to Fig.~2B and 2C in the main article illustrate the energy level structures and 
the relevant transition times for the DE and for the HH, respectively. Using 
these rates in Eq.~\ref{eq:Ind_model} results in indistinguishability of about 20\% for 
the DE-cluster and 95\% for the HH-Trion photons. These estimates agree with the 
experimental results of 17\% and 95\%  for the DE and HH, respectively, presented in the main article. 

\section*{SM5. The $X^{+1}$ optical transitions in the presence of an externally applied in-plane magnetic field.}

\begin{figure}
\begin{centering}
\includegraphics[width=0.7\columnwidth]{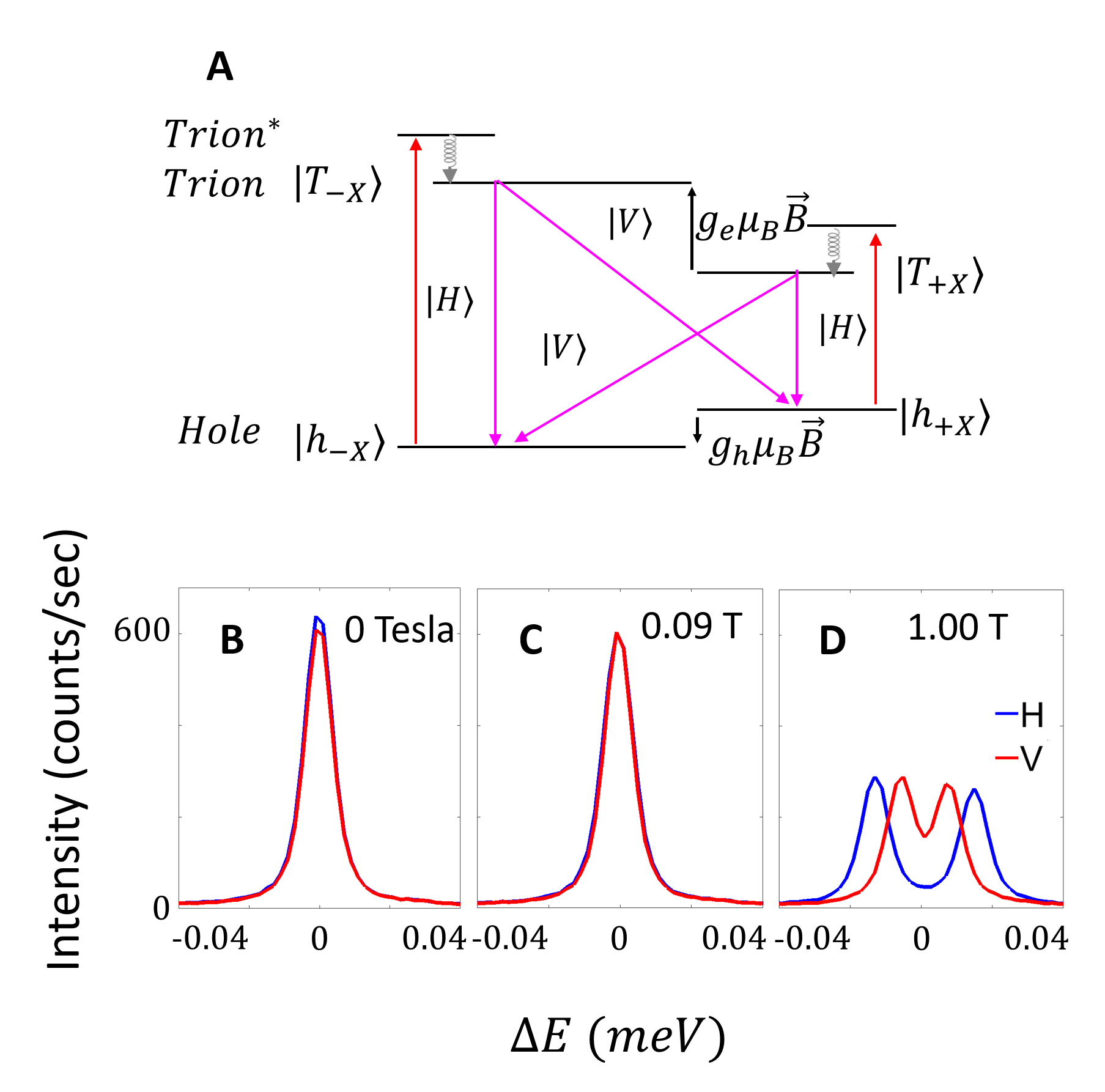}
\par\end{centering}
\caption{{\bf Spectroscopy of the positive trion ($X^{+1}$) in the presence of externally applied in-plane magnetic field.} ({\bf A}) Schematics of the Trion-HH energy levels and optical transitions in the presence of an in-plane external magnetic field. Here H=X (V=Y) describes horizontally (vertically) rectilinearly polarized optical transition. 
({\bf B-D}) Polarization-sensitive PL spectral measurements of the Trion-HH transition for various magnetic-fields.}
\end{figure}
The externally applied magnetic field reduces the indistinguishability of the HH-cluster photons.
In Fig.~S3A, we schematically describe the HH and positive-trion energy levels and the selection rules 
for optical transitions between these levels in the presence of an in-plane magnetic field. The field 
lifts the degeneracy of both the HH and the trion qubits. The resulted HH precession is used for 
implementing the Hadamard gate on the spin. The precession of the trion is used for tomography of the spin qubit \cite{Cogan2020}. 
In Fig.~S3B-D, we show polarization-sensitive spectral 
measurements of the $X^{+1}$ spectral line for various magnetic fields. Without a magnetic field, 
the line is completely degenerate. For a small field of 0.09 Tesla, the degeneracy removal results in a 
small broadening of the spectral line. This degeneracy removal is more than enough, however, for our 
purpose, as shown in the main text. For a  field of 1.00 Tesla, the spectral line is split into four 
components, rectilinearly cross-polarized, in perfect agreement with the scheme of Fig.~S3A.

Fig.~2D in the main article presents indistinguishability measurements of the HH-cluster photons 
for various magnetic fields. As the field increases, the indistinguishability decreases due to the 
field-induced spectral broadening. At large fields, when the Zeeman splitting surpasses the radiative 
linewidth, quantum beats are observed in the time-dependent correlation measurements \cite{Legero2004}. 
The quantum beats oscillation frequency exactly matches the energy difference between the V-polarized 
components of the $X^{+1}$  line in Fig. S3D.


\input{cluster_sm.bbl}

\bibliographystyle{aipnum4-1}

%% file: cluster_ms.bbl
%

%% file: cluster_sm.bbl
%

%% file: cluster_ms.bbl
\begin{thebibliography}{36}%
\makeatletter
\providecommand \@ifxundefined [1]{%
 \@ifx{#1\undefined}
}%
\providecommand \@ifnum [1]{%
 \ifnum #1\expandafter \@firstoftwo
 \else \expandafter \@secondoftwo
 \fi
}%
\providecommand \@ifx [1]{%
 \ifx #1\expandafter \@firstoftwo
 \else \expandafter \@secondoftwo
 \fi
}%
\providecommand \natexlab [1]{#1}%
\providecommand \enquote  [1]{``#1''}%
\providecommand \bibnamefont  [1]{#1}%
\providecommand \bibfnamefont [1]{#1}%
\providecommand \citenamefont [1]{#1}%
\providecommand \href@noop [0]{\@secondoftwo}%
\providecommand \href [0]{\begingroup \@sanitize@url \@href}%
\providecommand \@href[1]{\@@startlink{#1}\@@href}%
\providecommand \@@href[1]{\endgroup#1\@@endlink}%
\providecommand \@sanitize@url [0]{\catcode `\\12\catcode `\$12\catcode
  `\&12\catcode `\#12\catcode `\^12\catcode `\_12\catcode `\%12\relax}%
\providecommand \@@startlink[1]{}%
\providecommand \@@endlink[0]{}%
\providecommand \url  [0]{\begingroup\@sanitize@url \@url }%
\providecommand \@url [1]{\endgroup\@href {#1}{\urlprefix }}%
\providecommand \urlprefix  [0]{URL }%
\providecommand \Eprint [0]{\href }%
\providecommand \doibase [0]{http://dx.doi.org/}%
\providecommand \selectlanguage [0]{\@gobble}%
\providecommand \bibinfo  [0]{\@secondoftwo}%
\providecommand \bibfield  [0]{\@secondoftwo}%
\providecommand \translation [1]{[#1]}%
\providecommand \BibitemOpen [0]{}%
\providecommand \bibitemStop [0]{}%
\providecommand \bibitemNoStop [0]{.\EOS\space}%
\providecommand \EOS [0]{\spacefactor3000\relax}%
\providecommand \BibitemShut  [1]{\csname bibitem#1\endcsname}%
\let\auto@bib@innerbib\@empty
\bibitem [{\citenamefont {Briegel}\ \emph {et~al.}(1998)\citenamefont
  {Briegel}, \citenamefont {D{\"u}r}, \citenamefont {Cirac},\ and\
  \citenamefont {Zoller}}]{Briegel1998}%
  \BibitemOpen
  \bibfield  {author} {\bibinfo {author} {\bibfnamefont {H.-J.}\ \bibnamefont
  {Briegel}}, \bibinfo {author} {\bibfnamefont {W.}~\bibnamefont {D{\"u}r}},
  \bibinfo {author} {\bibfnamefont {J.~I.}\ \bibnamefont {Cirac}}, \ and\
  \bibinfo {author} {\bibfnamefont {P.}~\bibnamefont {Zoller}},\ }\href
  {\doibase 10.1103/physrevlett.81.5932} {\bibfield  {journal} {\bibinfo
  {journal} {Physical Review Letters}\ }\textbf {\bibinfo {volume} {81}},\
  \bibinfo {pages} {5932} (\bibinfo {year} {1998})}\BibitemShut {NoStop}%
\bibitem [{\citenamefont {Duan}\ \emph {et~al.}(2001)\citenamefont {Duan},
  \citenamefont {Lukin}, \citenamefont {Cirac},\ and\ \citenamefont
  {Zoller}}]{Duan2001}%
  \BibitemOpen
  \bibfield  {author} {\bibinfo {author} {\bibfnamefont {L.-M.}\ \bibnamefont
  {Duan}}, \bibinfo {author} {\bibfnamefont {M.~D.}\ \bibnamefont {Lukin}},
  \bibinfo {author} {\bibfnamefont {J.~I.}\ \bibnamefont {Cirac}}, \ and\
  \bibinfo {author} {\bibfnamefont {P.}~\bibnamefont {Zoller}},\ }\href
  {\doibase 10.1038/35106500} {\bibfield  {journal} {\bibinfo  {journal}
  {Nature}\ }\textbf {\bibinfo {volume} {414}},\ \bibinfo {pages} {413}
  (\bibinfo {year} {2001})}\BibitemShut {NoStop}%
\bibitem [{\citenamefont {Wehner}, \citenamefont {Elkouss},\ and\ \citenamefont
  {Hanson}(2018)}]{Wehner2018}%
  \BibitemOpen
  \bibfield  {author} {\bibinfo {author} {\bibfnamefont {S.}~\bibnamefont
  {Wehner}}, \bibinfo {author} {\bibfnamefont {D.}~\bibnamefont {Elkouss}}, \
  and\ \bibinfo {author} {\bibfnamefont {R.}~\bibnamefont {Hanson}},\ }\href
  {\doibase 10.1126/science.aam9288} {\bibfield  {journal} {\bibinfo  {journal}
  {Science}\ }\textbf {\bibinfo {volume} {362}},\ \bibinfo {pages} {303}
  (\bibinfo {year} {2018})}\BibitemShut {NoStop}%
\bibitem [{\citenamefont {Raussendorf}\ and\ \citenamefont
  {Briegel}(2001)}]{Raussendorf2001}%
  \BibitemOpen
  \bibfield  {author} {\bibinfo {author} {\bibfnamefont {R.}~\bibnamefont
  {Raussendorf}}\ and\ \bibinfo {author} {\bibfnamefont {H.~J.}\ \bibnamefont
  {Briegel}},\ }\href {\doibase 10.1103/physrevlett.86.5188} {\bibfield
  {journal} {\bibinfo  {journal} {Physical Review Letters}\ }\textbf {\bibinfo
  {volume} {86}},\ \bibinfo {pages} {5188} (\bibinfo {year}
  {2001})}\BibitemShut {NoStop}%
\bibitem [{\citenamefont {Zwerger}, \citenamefont {D{\"u}r},\ and\
  \citenamefont {Briegel}(2012)}]{Zwerger2012}%
  \BibitemOpen
  \bibfield  {author} {\bibinfo {author} {\bibfnamefont {M.}~\bibnamefont
  {Zwerger}}, \bibinfo {author} {\bibfnamefont {W.}~\bibnamefont {D{\"u}r}}, \
  and\ \bibinfo {author} {\bibfnamefont {H.~J.}\ \bibnamefont {Briegel}},\
  }\href {\doibase 10.1103/physreva.85.062326} {\bibfield  {journal} {\bibinfo
  {journal} {Physical Review A}\ }\textbf {\bibinfo {volume} {85}},\ \bibinfo
  {pages} {062326} (\bibinfo {year} {2012})}\BibitemShut {NoStop}%
\bibitem [{\citenamefont {Hein}, \citenamefont {Eisert},\ and\ \citenamefont
  {Briegel}(2004)}]{Hein2004}%
  \BibitemOpen
  \bibfield  {author} {\bibinfo {author} {\bibfnamefont {M.}~\bibnamefont
  {Hein}}, \bibinfo {author} {\bibfnamefont {J.}~\bibnamefont {Eisert}}, \ and\
  \bibinfo {author} {\bibfnamefont {H.~J.}\ \bibnamefont {Briegel}},\ }\href
  {\doibase 10.1103/physreva.69.062311} {\bibfield  {journal} {\bibinfo
  {journal} {Physical Review A}\ }\textbf {\bibinfo {volume} {69}},\ \bibinfo
  {pages} {062311} (\bibinfo {year} {2004})}\BibitemShut {NoStop}%
\bibitem [{\citenamefont {Buterakos}, \citenamefont {Barnes},\ and\
  \citenamefont {Economou}(2017)}]{Buterakos2017}%
  \BibitemOpen
  \bibfield  {author} {\bibinfo {author} {\bibfnamefont {D.}~\bibnamefont
  {Buterakos}}, \bibinfo {author} {\bibfnamefont {E.}~\bibnamefont {Barnes}}, \
  and\ \bibinfo {author} {\bibfnamefont {S.~E.}\ \bibnamefont {Economou}},\
  }\href {\doibase 10.1103/physrevx.7.041023} {\bibfield  {journal} {\bibinfo
  {journal} {Physical Review X}\ }\textbf {\bibinfo {volume} {7}},\ \bibinfo
  {pages} {041023} (\bibinfo {year} {2017})}\BibitemShut {NoStop}%
\bibitem [{\citenamefont {Azuma}, \citenamefont {Tamaki},\ and\ \citenamefont
  {Lo}(2015)}]{Azuma2015}%
  \BibitemOpen
  \bibfield  {author} {\bibinfo {author} {\bibfnamefont {K.}~\bibnamefont
  {Azuma}}, \bibinfo {author} {\bibfnamefont {K.}~\bibnamefont {Tamaki}}, \
  and\ \bibinfo {author} {\bibfnamefont {H.-K.}\ \bibnamefont {Lo}},\ }\href
  {\doibase 10.1038/ncomms7787} {\bibfield  {journal} {\bibinfo  {journal}
  {Nature Communications}\ }\textbf {\bibinfo {volume} {6}},\ \bibinfo {pages}
  {6787} (\bibinfo {year} {2015})}\BibitemShut {NoStop}%
\bibitem [{\citenamefont {Munro}\ \emph {et~al.}(2012)\citenamefont {Munro},
  \citenamefont {Stephens}, \citenamefont {Devitt}, \citenamefont {Harrison},\
  and\ \citenamefont {Nemoto}}]{Munro2012}%
  \BibitemOpen
  \bibfield  {author} {\bibinfo {author} {\bibfnamefont {W.~J.}\ \bibnamefont
  {Munro}}, \bibinfo {author} {\bibfnamefont {A.~M.}\ \bibnamefont {Stephens}},
  \bibinfo {author} {\bibfnamefont {S.~J.}\ \bibnamefont {Devitt}}, \bibinfo
  {author} {\bibfnamefont {K.~A.}\ \bibnamefont {Harrison}}, \ and\ \bibinfo
  {author} {\bibfnamefont {K.}~\bibnamefont {Nemoto}},\ }\href {\doibase
  10.1038/nphoton.2012.243} {\bibfield  {journal} {\bibinfo  {journal} {Nature
  Photonics}\ }\textbf {\bibinfo {volume} {6}},\ \bibinfo {pages} {777}
  (\bibinfo {year} {2012})}\BibitemShut {NoStop}%
\bibitem [{\citenamefont {Larsen}\ \emph {et~al.}(2019)\citenamefont {Larsen},
  \citenamefont {Guo}, \citenamefont {Breum}, \citenamefont
  {Neergaard-Nielsen},\ and\ \citenamefont {Andersen}}]{Larsen2019}%
  \BibitemOpen
  \bibfield  {author} {\bibinfo {author} {\bibfnamefont {M.~V.}\ \bibnamefont
  {Larsen}}, \bibinfo {author} {\bibfnamefont {X.}~\bibnamefont {Guo}},
  \bibinfo {author} {\bibfnamefont {C.~R.}\ \bibnamefont {Breum}}, \bibinfo
  {author} {\bibfnamefont {J.~S.}\ \bibnamefont {Neergaard-Nielsen}}, \ and\
  \bibinfo {author} {\bibfnamefont {U.~L.}\ \bibnamefont {Andersen}},\ }\href
  {\doibase 10.1126/science.aay4354} {\bibfield  {journal} {\bibinfo  {journal}
  {Science}\ }\textbf {\bibinfo {volume} {366}},\ \bibinfo {pages} {369}
  (\bibinfo {year} {2019})}\BibitemShut {NoStop}%
\bibitem [{\citenamefont {Li}\ \emph {et~al.}(2020)\citenamefont {Li},
  \citenamefont {Qin}, \citenamefont {Chen}, \citenamefont {Duan},
  \citenamefont {Yu}, \citenamefont {Huo}, \citenamefont {Höfling},
  \citenamefont {Lu}, \citenamefont {Chen},\ and\ \citenamefont
  {Pan}}]{Li2020}%
  \BibitemOpen
  \bibfield  {author} {\bibinfo {author} {\bibfnamefont {J.-P.}\ \bibnamefont
  {Li}}, \bibinfo {author} {\bibfnamefont {J.}~\bibnamefont {Qin}}, \bibinfo
  {author} {\bibfnamefont {A.}~\bibnamefont {Chen}}, \bibinfo {author}
  {\bibfnamefont {Z.-C.}\ \bibnamefont {Duan}}, \bibinfo {author}
  {\bibfnamefont {Y.}~\bibnamefont {Yu}}, \bibinfo {author} {\bibfnamefont
  {Y.}~\bibnamefont {Huo}}, \bibinfo {author} {\bibfnamefont {S.}~\bibnamefont
  {Höfling}}, \bibinfo {author} {\bibfnamefont {C.-Y.}\ \bibnamefont {Lu}},
  \bibinfo {author} {\bibfnamefont {K.}~\bibnamefont {Chen}}, \ and\ \bibinfo
  {author} {\bibfnamefont {J.-W.}\ \bibnamefont {Pan}},\ }\href {\doibase
  10.1021/acsphotonics.0c00192} {\bibfield  {journal} {\bibinfo  {journal}
  {{ACS} Photonics}\ }\textbf {\bibinfo {volume} {7}},\ \bibinfo {pages} {1603}
  (\bibinfo {year} {2020})}\BibitemShut {NoStop}%
\bibitem [{\citenamefont {Istrati}\ \emph {et~al.}(2020)\citenamefont
  {Istrati}, \citenamefont {Pilnyak}, \citenamefont {Loredo}, \citenamefont
  {Ant{\'{o}}n}, \citenamefont {Somaschi}, \citenamefont {Hilaire},
  \citenamefont {Ollivier}, \citenamefont {Esmann}, \citenamefont {Cohen},
  \citenamefont {Vidro}, \citenamefont {Millet}, \citenamefont
  {Lema{\^{\i}}tre}, \citenamefont {Sagnes}, \citenamefont {Harouri},
  \citenamefont {Lanco}, \citenamefont {Senellart},\ and\ \citenamefont
  {Eisenberg}}]{Istrati2020}%
  \BibitemOpen
  \bibfield  {author} {\bibinfo {author} {\bibfnamefont {D.}~\bibnamefont
  {Istrati}}, \bibinfo {author} {\bibfnamefont {Y.}~\bibnamefont {Pilnyak}},
  \bibinfo {author} {\bibfnamefont {J.~C.}\ \bibnamefont {Loredo}}, \bibinfo
  {author} {\bibfnamefont {C.}~\bibnamefont {Ant{\'{o}}n}}, \bibinfo {author}
  {\bibfnamefont {N.}~\bibnamefont {Somaschi}}, \bibinfo {author}
  {\bibfnamefont {P.}~\bibnamefont {Hilaire}}, \bibinfo {author} {\bibfnamefont
  {H.}~\bibnamefont {Ollivier}}, \bibinfo {author} {\bibfnamefont
  {M.}~\bibnamefont {Esmann}}, \bibinfo {author} {\bibfnamefont
  {L.}~\bibnamefont {Cohen}}, \bibinfo {author} {\bibfnamefont
  {L.}~\bibnamefont {Vidro}}, \bibinfo {author} {\bibfnamefont
  {C.}~\bibnamefont {Millet}}, \bibinfo {author} {\bibfnamefont
  {A.}~\bibnamefont {Lema{\^{\i}}tre}}, \bibinfo {author} {\bibfnamefont
  {I.}~\bibnamefont {Sagnes}}, \bibinfo {author} {\bibfnamefont
  {A.}~\bibnamefont {Harouri}}, \bibinfo {author} {\bibfnamefont
  {L.}~\bibnamefont {Lanco}}, \bibinfo {author} {\bibfnamefont
  {P.}~\bibnamefont {Senellart}}, \ and\ \bibinfo {author} {\bibfnamefont
  {H.~S.}\ \bibnamefont {Eisenberg}},\ }\href {\doibase
  10.1038/s41467-020-19341-4} {\bibfield  {journal} {\bibinfo  {journal}
  {Nature Communications}\ }\textbf {\bibinfo {volume} {11}},\ \bibinfo {pages}
  {5501} (\bibinfo {year} {2020})}\BibitemShut {NoStop}%
\bibitem [{\citenamefont {Besse}\ \emph {et~al.}(2020)\citenamefont {Besse},
  \citenamefont {Reuer}, \citenamefont {Collodo}, \citenamefont {Wulff},
  \citenamefont {Wernli}, \citenamefont {Copetudo}, \citenamefont {Malz},
  \citenamefont {Magnard}, \citenamefont {Akin}, \citenamefont {Gabureac},
  \citenamefont {Norris}, \citenamefont {Cirac}, \citenamefont {Wallraff},\
  and\ \citenamefont {Eichler}}]{Besse2020}%
  \BibitemOpen
  \bibfield  {author} {\bibinfo {author} {\bibfnamefont {J.-C.}\ \bibnamefont
  {Besse}}, \bibinfo {author} {\bibfnamefont {K.}~\bibnamefont {Reuer}},
  \bibinfo {author} {\bibfnamefont {M.~C.}\ \bibnamefont {Collodo}}, \bibinfo
  {author} {\bibfnamefont {A.}~\bibnamefont {Wulff}}, \bibinfo {author}
  {\bibfnamefont {L.}~\bibnamefont {Wernli}}, \bibinfo {author} {\bibfnamefont
  {A.}~\bibnamefont {Copetudo}}, \bibinfo {author} {\bibfnamefont
  {D.}~\bibnamefont {Malz}}, \bibinfo {author} {\bibfnamefont {P.}~\bibnamefont
  {Magnard}}, \bibinfo {author} {\bibfnamefont {A.}~\bibnamefont {Akin}},
  \bibinfo {author} {\bibfnamefont {M.}~\bibnamefont {Gabureac}}, \bibinfo
  {author} {\bibfnamefont {G.~J.}\ \bibnamefont {Norris}}, \bibinfo {author}
  {\bibfnamefont {J.~I.}\ \bibnamefont {Cirac}}, \bibinfo {author}
  {\bibfnamefont {A.}~\bibnamefont {Wallraff}}, \ and\ \bibinfo {author}
  {\bibfnamefont {C.}~\bibnamefont {Eichler}},\ }\href {\doibase
  10.1038/s41467-020-18635-x} {\bibfield  {journal} {\bibinfo  {journal}
  {Nature Communications}\ }\textbf {\bibinfo {volume} {11}},\ \bibinfo {pages}
  {4877} (\bibinfo {year} {2020})}\BibitemShut {NoStop}%
\bibitem [{\citenamefont {Aspect}, \citenamefont {Grangier},\ and\
  \citenamefont {Roger}(1982)}]{Aspect1982}%
  \BibitemOpen
  \bibfield  {author} {\bibinfo {author} {\bibfnamefont {A.}~\bibnamefont
  {Aspect}}, \bibinfo {author} {\bibfnamefont {P.}~\bibnamefont {Grangier}}, \
  and\ \bibinfo {author} {\bibfnamefont {G.}~\bibnamefont {Roger}},\ }\href
  {\doibase 10.1103/physrevlett.49.91} {\bibfield  {journal} {\bibinfo
  {journal} {Physical Review Letters}\ }\textbf {\bibinfo {volume} {49}},\
  \bibinfo {pages} {91} (\bibinfo {year} {1982})}\BibitemShut {NoStop}%
\bibitem [{\citenamefont {Akopian}\ \emph {et~al.}(2006)\citenamefont
  {Akopian}, \citenamefont {Lindner}, \citenamefont {Poem}, \citenamefont
  {Berlatzky}, \citenamefont {Avron}, \citenamefont {Gershoni}, \citenamefont
  {Gerardot},\ and\ \citenamefont {Petroff}}]{Akopian2006}%
  \BibitemOpen
  \bibfield  {author} {\bibinfo {author} {\bibfnamefont {N.}~\bibnamefont
  {Akopian}}, \bibinfo {author} {\bibfnamefont {N.~H.}\ \bibnamefont
  {Lindner}}, \bibinfo {author} {\bibfnamefont {E.}~\bibnamefont {Poem}},
  \bibinfo {author} {\bibfnamefont {Y.}~\bibnamefont {Berlatzky}}, \bibinfo
  {author} {\bibfnamefont {J.}~\bibnamefont {Avron}}, \bibinfo {author}
  {\bibfnamefont {D.}~\bibnamefont {Gershoni}}, \bibinfo {author}
  {\bibfnamefont {B.~D.}\ \bibnamefont {Gerardot}}, \ and\ \bibinfo {author}
  {\bibfnamefont {P.~M.}\ \bibnamefont {Petroff}},\ }\href {\doibase
  10.1103/physrevlett.96.130501} {\bibfield  {journal} {\bibinfo  {journal}
  {Physical Review Letters}\ }\textbf {\bibinfo {volume} {96}},\ \bibinfo
  {pages} {130501} (\bibinfo {year} {2006})}\BibitemShut {NoStop}%
\bibitem [{\citenamefont {Senellart}, \citenamefont {Solomon},\ and\
  \citenamefont {White}(2017)}]{Senellart2017}%
  \BibitemOpen
  \bibfield  {author} {\bibinfo {author} {\bibfnamefont {P.}~\bibnamefont
  {Senellart}}, \bibinfo {author} {\bibfnamefont {G.}~\bibnamefont {Solomon}},
  \ and\ \bibinfo {author} {\bibfnamefont {A.}~\bibnamefont {White}},\ }\href
  {\doibase 10.1038/nnano.2017.218} {\bibfield  {journal} {\bibinfo  {journal}
  {Nature Nanotechnology}\ }\textbf {\bibinfo {volume} {12}},\ \bibinfo {pages}
  {1026} (\bibinfo {year} {2017})}\BibitemShut {NoStop}%
\bibitem [{\citenamefont {Appel}\ \emph {et~al.}(2021)\citenamefont {Appel},
  \citenamefont {Tiranov}, \citenamefont {Javadi}, \citenamefont {Löbl},
  \citenamefont {Wang}, \citenamefont {Scholz}, \citenamefont {Wieck},
  \citenamefont {Ludwig}, \citenamefont {Warburton},\ and\ \citenamefont
  {Lodahl}}]{Appel2021}%
  \BibitemOpen
  \bibfield  {author} {\bibinfo {author} {\bibfnamefont {M.~H.}\ \bibnamefont
  {Appel}}, \bibinfo {author} {\bibfnamefont {A.}~\bibnamefont {Tiranov}},
  \bibinfo {author} {\bibfnamefont {A.}~\bibnamefont {Javadi}}, \bibinfo
  {author} {\bibfnamefont {M.~C.}\ \bibnamefont {Löbl}}, \bibinfo {author}
  {\bibfnamefont {Y.}~\bibnamefont {Wang}}, \bibinfo {author} {\bibfnamefont
  {S.}~\bibnamefont {Scholz}}, \bibinfo {author} {\bibfnamefont {A.~D.}\
  \bibnamefont {Wieck}}, \bibinfo {author} {\bibfnamefont {A.}~\bibnamefont
  {Ludwig}}, \bibinfo {author} {\bibfnamefont {R.~J.}\ \bibnamefont
  {Warburton}}, \ and\ \bibinfo {author} {\bibfnamefont {P.}~\bibnamefont
  {Lodahl}},\ }\href {\doibase 10.1103/physrevlett.126.013602} {\bibfield
  {journal} {\bibinfo  {journal} {Physical Review Letters}\ }\textbf {\bibinfo
  {volume} {126}},\ \bibinfo {pages} {013602} (\bibinfo {year}
  {2021})}\BibitemShut {NoStop}%
\bibitem [{\citenamefont {Lindner}\ and\ \citenamefont
  {Rudolph}(2009)}]{Lindner2009}%
  \BibitemOpen
  \bibfield  {author} {\bibinfo {author} {\bibfnamefont {N.~H.}\ \bibnamefont
  {Lindner}}\ and\ \bibinfo {author} {\bibfnamefont {T.}~\bibnamefont
  {Rudolph}},\ }\href {\doibase 10.1103/physrevlett.103.113602} {\bibfield
  {journal} {\bibinfo  {journal} {Physical Review Letters}\ }\textbf {\bibinfo
  {volume} {103}},\ \bibinfo {pages} {113602} (\bibinfo {year}
  {2009})}\BibitemShut {NoStop}%
\bibitem [{\citenamefont {Schwartz}\ \emph {et~al.}(2016)\citenamefont
  {Schwartz}, \citenamefont {Cogan}, \citenamefont {Schmidgall}, \citenamefont
  {Don}, \citenamefont {Gantz}, \citenamefont {Kenneth}, \citenamefont
  {Lindner},\ and\ \citenamefont {Gershoni}}]{Schwartz2016}%
  \BibitemOpen
  \bibfield  {author} {\bibinfo {author} {\bibfnamefont {I.}~\bibnamefont
  {Schwartz}}, \bibinfo {author} {\bibfnamefont {D.}~\bibnamefont {Cogan}},
  \bibinfo {author} {\bibfnamefont {E.~R.}\ \bibnamefont {Schmidgall}},
  \bibinfo {author} {\bibfnamefont {Y.}~\bibnamefont {Don}}, \bibinfo {author}
  {\bibfnamefont {L.}~\bibnamefont {Gantz}}, \bibinfo {author} {\bibfnamefont
  {O.}~\bibnamefont {Kenneth}}, \bibinfo {author} {\bibfnamefont {N.~H.}\
  \bibnamefont {Lindner}}, \ and\ \bibinfo {author} {\bibfnamefont
  {D.}~\bibnamefont {Gershoni}},\ }\href {\doibase 10.1126/science.aah4758}
  {\bibfield  {journal} {\bibinfo  {journal} {Science}\ }\textbf {\bibinfo
  {volume} {354}},\ \bibinfo {pages} {434} (\bibinfo {year}
  {2016})}\BibitemShut {NoStop}%
\bibitem [{\citenamefont {Cogan}\ \emph {et~al.}(2018)\citenamefont {Cogan},
  \citenamefont {Kenneth}, \citenamefont {Lindner}, \citenamefont {Peniakov},
  \citenamefont {Hopfmann}, \citenamefont {Dalacu}, \citenamefont {Poole},
  \citenamefont {Hawrylak},\ and\ \citenamefont {Gershoni}}]{Cogan2018}%
  \BibitemOpen
  \bibfield  {author} {\bibinfo {author} {\bibfnamefont {D.}~\bibnamefont
  {Cogan}}, \bibinfo {author} {\bibfnamefont {O.}~\bibnamefont {Kenneth}},
  \bibinfo {author} {\bibfnamefont {N.~H.}\ \bibnamefont {Lindner}}, \bibinfo
  {author} {\bibfnamefont {G.}~\bibnamefont {Peniakov}}, \bibinfo {author}
  {\bibfnamefont {C.}~\bibnamefont {Hopfmann}}, \bibinfo {author}
  {\bibfnamefont {D.}~\bibnamefont {Dalacu}}, \bibinfo {author} {\bibfnamefont
  {P.~J.}\ \bibnamefont {Poole}}, \bibinfo {author} {\bibfnamefont
  {P.}~\bibnamefont {Hawrylak}}, \ and\ \bibinfo {author} {\bibfnamefont
  {D.}~\bibnamefont {Gershoni}},\ }\href {\doibase 10.1103/physrevx.8.041050}
  {\bibfield  {journal} {\bibinfo  {journal} {Physical Review X}\ }\textbf
  {\bibinfo {volume} {8}},\ \bibinfo {pages} {041050} (\bibinfo {year}
  {2018})}\BibitemShut {NoStop}%
\bibitem [{\citenamefont {Bechtold}\ \emph {et~al.}(2015)\citenamefont
  {Bechtold}, \citenamefont {Rauch}, \citenamefont {Li}, \citenamefont
  {Simmet}, \citenamefont {Ardelt}, \citenamefont {Regler}, \citenamefont
  {Muller}, \citenamefont {Sinitsyn},\ and\ \citenamefont
  {Finley}}]{Bechtold2015}%
  \BibitemOpen
  \bibfield  {author} {\bibinfo {author} {\bibfnamefont {A.}~\bibnamefont
  {Bechtold}}, \bibinfo {author} {\bibfnamefont {D.}~\bibnamefont {Rauch}},
  \bibinfo {author} {\bibfnamefont {F.}~\bibnamefont {Li}}, \bibinfo {author}
  {\bibfnamefont {T.}~\bibnamefont {Simmet}}, \bibinfo {author} {\bibfnamefont
  {P.-L.}\ \bibnamefont {Ardelt}}, \bibinfo {author} {\bibfnamefont
  {A.}~\bibnamefont {Regler}}, \bibinfo {author} {\bibfnamefont
  {K.}~\bibnamefont {Muller}}, \bibinfo {author} {\bibfnamefont {N.~A.}\
  \bibnamefont {Sinitsyn}}, \ and\ \bibinfo {author} {\bibfnamefont {J.~J.}\
  \bibnamefont {Finley}},\ }\href {http://dx.doi.org/10.1038/nphys3470}
  {\bibfield  {journal} {\bibinfo  {journal} {Nature Physics}\ }\textbf
  {\bibinfo {volume} {11}},\ \bibinfo {pages} {1005} (\bibinfo {year}
  {2015})}\BibitemShut {NoStop}%
\bibitem [{\citenamefont {Kiraz}, \citenamefont {Atat{\"u}re},\ and\
  \citenamefont {Imamo{\u{g}}lu}(2004)}]{Kiraz2004}%
  \BibitemOpen
  \bibfield  {author} {\bibinfo {author} {\bibfnamefont {A.}~\bibnamefont
  {Kiraz}}, \bibinfo {author} {\bibfnamefont {M.}~\bibnamefont {Atat{\"u}re}},
  \ and\ \bibinfo {author} {\bibfnamefont {A.}~\bibnamefont {Imamo{\u{g}}lu}},\
  }\href {\doibase 10.1103/physreva.69.032305} {\bibfield  {journal} {\bibinfo
  {journal} {Physical Review A}\ }\textbf {\bibinfo {volume} {69}},\ \bibinfo
  {pages} {032305} (\bibinfo {year} {2004})}\BibitemShut {NoStop}%
\bibitem [{\citenamefont {Schwartz}\ \emph {et~al.}(2015)\citenamefont
  {Schwartz}, \citenamefont {Schmidgall}, \citenamefont {Gantz}, \citenamefont
  {Cogan}, \citenamefont {Bordo}, \citenamefont {Don}, \citenamefont
  {Zielinski},\ and\ \citenamefont {Gershoni}}]{Schwartz2015}%
  \BibitemOpen
  \bibfield  {author} {\bibinfo {author} {\bibfnamefont {I.}~\bibnamefont
  {Schwartz}}, \bibinfo {author} {\bibfnamefont {E.~R.}\ \bibnamefont
  {Schmidgall}}, \bibinfo {author} {\bibfnamefont {L.}~\bibnamefont {Gantz}},
  \bibinfo {author} {\bibfnamefont {D.}~\bibnamefont {Cogan}}, \bibinfo
  {author} {\bibfnamefont {E.}~\bibnamefont {Bordo}}, \bibinfo {author}
  {\bibfnamefont {Y.}~\bibnamefont {Don}}, \bibinfo {author} {\bibfnamefont
  {M.}~\bibnamefont {Zielinski}}, \ and\ \bibinfo {author} {\bibfnamefont
  {D.}~\bibnamefont {Gershoni}},\ }\href {\doibase 10.1103/physrevx.5.011009}
  {\bibfield  {journal} {\bibinfo  {journal} {Physical Review X}\ }\textbf
  {\bibinfo {volume} {5}},\ \bibinfo {pages} {011009} (\bibinfo {year}
  {2015})}\BibitemShut {NoStop}%
\bibitem [{\citenamefont {Gerardot}\ \emph {et~al.}(2008)\citenamefont
  {Gerardot}, \citenamefont {Brunner}, \citenamefont {Dalgarno}, \citenamefont
  {{\"O}hberg}, \citenamefont {Seidl}, \citenamefont {Kroner}, \citenamefont
  {Karrai}, \citenamefont {Stoltz}, \citenamefont {Petroff},\ and\
  \citenamefont {Warburton}}]{Gerardot2008}%
  \BibitemOpen
  \bibfield  {author} {\bibinfo {author} {\bibfnamefont {B.~D.}\ \bibnamefont
  {Gerardot}}, \bibinfo {author} {\bibfnamefont {D.}~\bibnamefont {Brunner}},
  \bibinfo {author} {\bibfnamefont {P.~A.}\ \bibnamefont {Dalgarno}}, \bibinfo
  {author} {\bibfnamefont {P.}~\bibnamefont {{\"O}hberg}}, \bibinfo {author}
  {\bibfnamefont {S.}~\bibnamefont {Seidl}}, \bibinfo {author} {\bibfnamefont
  {M.}~\bibnamefont {Kroner}}, \bibinfo {author} {\bibfnamefont
  {K.}~\bibnamefont {Karrai}}, \bibinfo {author} {\bibfnamefont {N.~G.}\
  \bibnamefont {Stoltz}}, \bibinfo {author} {\bibfnamefont {P.~M.}\
  \bibnamefont {Petroff}}, \ and\ \bibinfo {author} {\bibfnamefont {R.~J.}\
  \bibnamefont {Warburton}},\ }\href {\doibase 10.1038/nature06472} {\bibfield
  {journal} {\bibinfo  {journal} {Nature}\ }\textbf {\bibinfo {volume} {451}},\
  \bibinfo {pages} {441} (\bibinfo {year} {2008})}\BibitemShut {NoStop}%
\bibitem [{\citenamefont {Brunner}\ \emph {et~al.}(2009)\citenamefont
  {Brunner}, \citenamefont {Gerardot}, \citenamefont {Dalgarno}, \citenamefont
  {W{\"u}st}, \citenamefont {Karrai}, \citenamefont {Stoltz}, \citenamefont
  {Petroff},\ and\ \citenamefont {Warburton}}]{Brunner2009}%
  \BibitemOpen
  \bibfield  {author} {\bibinfo {author} {\bibfnamefont {D.}~\bibnamefont
  {Brunner}}, \bibinfo {author} {\bibfnamefont {B.~D.}\ \bibnamefont
  {Gerardot}}, \bibinfo {author} {\bibfnamefont {P.~A.}\ \bibnamefont
  {Dalgarno}}, \bibinfo {author} {\bibfnamefont {G.}~\bibnamefont {W{\"u}st}},
  \bibinfo {author} {\bibfnamefont {K.}~\bibnamefont {Karrai}}, \bibinfo
  {author} {\bibfnamefont {N.~G.}\ \bibnamefont {Stoltz}}, \bibinfo {author}
  {\bibfnamefont {P.~M.}\ \bibnamefont {Petroff}}, \ and\ \bibinfo {author}
  {\bibfnamefont {R.~J.}\ \bibnamefont {Warburton}},\ }\href {\doibase
  10.1126/science.1173684} {\bibfield  {journal} {\bibinfo  {journal}
  {Science}\ }\textbf {\bibinfo {volume} {325}},\ \bibinfo {pages} {70}
  (\bibinfo {year} {2009})}\BibitemShut {NoStop}%
\bibitem [{\citenamefont {Cogan}\ \emph {et~al.}(2020)\citenamefont {Cogan},
  \citenamefont {Peniakov}, \citenamefont {Su},\ and\ \citenamefont
  {Gershoni}}]{Cogan2020}%
  \BibitemOpen
  \bibfield  {author} {\bibinfo {author} {\bibfnamefont {D.}~\bibnamefont
  {Cogan}}, \bibinfo {author} {\bibfnamefont {G.}~\bibnamefont {Peniakov}},
  \bibinfo {author} {\bibfnamefont {Z.-E.}\ \bibnamefont {Su}}, \ and\ \bibinfo
  {author} {\bibfnamefont {D.}~\bibnamefont {Gershoni}},\ }\href {\doibase
  10.1103/physrevb.101.035424} {\bibfield  {journal} {\bibinfo  {journal}
  {Physical Review B}\ }\textbf {\bibinfo {volume} {101}},\ \bibinfo {pages}
  {035424} (\bibinfo {year} {2020})}\BibitemShut {NoStop}%
\bibitem [{\citenamefont {Benny}\ \emph {et~al.}(2012)\citenamefont {Benny},
  \citenamefont {Kodriano}, \citenamefont {Poem}, \citenamefont {Gershoni},
  \citenamefont {Truong},\ and\ \citenamefont {Petroff}}]{Benny2012}%
  \BibitemOpen
  \bibfield  {author} {\bibinfo {author} {\bibfnamefont {Y.}~\bibnamefont
  {Benny}}, \bibinfo {author} {\bibfnamefont {Y.}~\bibnamefont {Kodriano}},
  \bibinfo {author} {\bibfnamefont {E.}~\bibnamefont {Poem}}, \bibinfo {author}
  {\bibfnamefont {D.}~\bibnamefont {Gershoni}}, \bibinfo {author}
  {\bibfnamefont {T.~A.}\ \bibnamefont {Truong}}, \ and\ \bibinfo {author}
  {\bibfnamefont {P.~M.}\ \bibnamefont {Petroff}},\ }\href {\doibase
  10.1103/physrevb.86.085306} {\bibfield  {journal} {\bibinfo  {journal}
  {Physical Review B}\ }\textbf {\bibinfo {volume} {86}},\ \bibinfo {pages}
  {085306} (\bibinfo {year} {2012})}\BibitemShut {NoStop}%
\bibitem [{\citenamefont {Loss}\ and\ \citenamefont
  {DiVincenzo}(1998)}]{Loss1998}%
  \BibitemOpen
  \bibfield  {author} {\bibinfo {author} {\bibfnamefont {D.}~\bibnamefont
  {Loss}}\ and\ \bibinfo {author} {\bibfnamefont {D.~P.}\ \bibnamefont
  {DiVincenzo}},\ }\href {\doibase 10.1103/physreva.57.120} {\bibfield
  {journal} {\bibinfo  {journal} {Physical Review A}\ }\textbf {\bibinfo
  {volume} {57}},\ \bibinfo {pages} {120} (\bibinfo {year} {1998})}\BibitemShut
  {NoStop}%
\bibitem [{SM()}]{SM}%
  \BibitemOpen
  \href@noop {} {\enquote {\bibinfo {title} {See supplementary materials},}\
  }\BibitemShut {NoStop}%
\bibitem [{\citenamefont {Cogan}\ \emph {et~al.}(2021)\citenamefont {Cogan},
  \citenamefont {Su}, \citenamefont {Kenneth},\ and\ \citenamefont
  {Gershoni}}]{Cogan2021}%
  \BibitemOpen
  \bibfield  {author} {\bibinfo {author} {\bibfnamefont {D.}~\bibnamefont
  {Cogan}}, \bibinfo {author} {\bibfnamefont {Z.-E.}\ \bibnamefont {Su}},
  \bibinfo {author} {\bibfnamefont {O.}~\bibnamefont {Kenneth}}, \ and\
  \bibinfo {author} {\bibfnamefont {D.}~\bibnamefont {Gershoni}},\ }\href@noop
  {} {\bibfield  {journal} {\bibinfo  {journal} {arXiv:2108.05173}\ } (\bibinfo
  {year} {2021})},\ \Eprint
  {http://arxiv.org/abs/http://arxiv.org/abs/2108.05173v1}
  {http://arxiv.org/abs/2108.05173v1} \BibitemShut {NoStop}%
\bibitem [{\citenamefont {Hong}, \citenamefont {Ou},\ and\ \citenamefont
  {Mandel}(1987)}]{Hong1987}%
  \BibitemOpen
  \bibfield  {author} {\bibinfo {author} {\bibfnamefont {C.~K.}\ \bibnamefont
  {Hong}}, \bibinfo {author} {\bibfnamefont {Z.~Y.}\ \bibnamefont {Ou}}, \ and\
  \bibinfo {author} {\bibfnamefont {L.}~\bibnamefont {Mandel}},\ }\href
  {\doibase 10.1103/physrevlett.59.2044} {\bibfield  {journal} {\bibinfo
  {journal} {Physical Review Letters}\ }\textbf {\bibinfo {volume} {59}},\
  \bibinfo {pages} {2044} (\bibinfo {year} {1987})}\BibitemShut {NoStop}%
\bibitem [{\citenamefont {Santori}\ \emph {et~al.}(2002)\citenamefont
  {Santori}, \citenamefont {Fattal}, \citenamefont {Vu{\v{c}}kovi{\'{c}}},
  \citenamefont {Solomon},\ and\ \citenamefont {Yamamoto}}]{Santori2002}%
  \BibitemOpen
  \bibfield  {author} {\bibinfo {author} {\bibfnamefont {C.}~\bibnamefont
  {Santori}}, \bibinfo {author} {\bibfnamefont {D.}~\bibnamefont {Fattal}},
  \bibinfo {author} {\bibfnamefont {J.}~\bibnamefont {Vu{\v{c}}kovi{\'{c}}}},
  \bibinfo {author} {\bibfnamefont {G.~S.}\ \bibnamefont {Solomon}}, \ and\
  \bibinfo {author} {\bibfnamefont {Y.}~\bibnamefont {Yamamoto}},\ }\href
  {\doibase 10.1038/nature01086} {\bibfield  {journal} {\bibinfo  {journal}
  {Nature}\ }\textbf {\bibinfo {volume} {419}},\ \bibinfo {pages} {594}
  (\bibinfo {year} {2002})}\BibitemShut {NoStop}%
\bibitem [{\citenamefont {Legero}\ \emph {et~al.}(2004)\citenamefont {Legero},
  \citenamefont {Wilk}, \citenamefont {Hennrich}, \citenamefont {Rempe},\ and\
  \citenamefont {Kuhn}}]{Legero2004}%
  \BibitemOpen
  \bibfield  {author} {\bibinfo {author} {\bibfnamefont {T.}~\bibnamefont
  {Legero}}, \bibinfo {author} {\bibfnamefont {T.}~\bibnamefont {Wilk}},
  \bibinfo {author} {\bibfnamefont {M.}~\bibnamefont {Hennrich}}, \bibinfo
  {author} {\bibfnamefont {G.}~\bibnamefont {Rempe}}, \ and\ \bibinfo {author}
  {\bibfnamefont {A.}~\bibnamefont {Kuhn}},\ }\href {\doibase
  10.1103/physrevlett.93.070503} {\bibfield  {journal} {\bibinfo  {journal}
  {Physical Review Letters}\ }\textbf {\bibinfo {volume} {93}},\ \bibinfo
  {pages} {070503} (\bibinfo {year} {2004})}\BibitemShut {NoStop}%
\bibitem [{\citenamefont {Verstraete}, \citenamefont {Popp},\ and\
  \citenamefont {Cirac}(2004)}]{Verstraete2004}%
  \BibitemOpen
  \bibfield  {author} {\bibinfo {author} {\bibfnamefont {F.}~\bibnamefont
  {Verstraete}}, \bibinfo {author} {\bibfnamefont {M.}~\bibnamefont {Popp}}, \
  and\ \bibinfo {author} {\bibfnamefont {J.~I.}\ \bibnamefont {Cirac}},\ }\href
  {\doibase 10.1103/physrevlett.92.027901} {\bibfield  {journal} {\bibinfo
  {journal} {Physical Review Letters}\ }\textbf {\bibinfo {volume} {92}},\
  \bibinfo {pages} {027901} (\bibinfo {year} {2004})}\BibitemShut {NoStop}%
\bibitem [{\citenamefont {Liu}\ \emph {et~al.}(2018)\citenamefont {Liu},
  \citenamefont {Brash}, \citenamefont {O'Hara}, \citenamefont {Martins},
  \citenamefont {Phillips}, \citenamefont {Coles}, \citenamefont {Royall},
  \citenamefont {Clarke}, \citenamefont {Bentham}, \citenamefont {Prtljaga},
  \citenamefont {Itskevich}, \citenamefont {Wilson}, \citenamefont {Skolnick},\
  and\ \citenamefont {Fox}}]{Liu2018}%
  \BibitemOpen
  \bibfield  {author} {\bibinfo {author} {\bibfnamefont {F.}~\bibnamefont
  {Liu}}, \bibinfo {author} {\bibfnamefont {A.~J.}\ \bibnamefont {Brash}},
  \bibinfo {author} {\bibfnamefont {J.}~\bibnamefont {O'Hara}}, \bibinfo
  {author} {\bibfnamefont {L.~M. P.~P.}\ \bibnamefont {Martins}}, \bibinfo
  {author} {\bibfnamefont {C.~L.}\ \bibnamefont {Phillips}}, \bibinfo {author}
  {\bibfnamefont {R.~J.}\ \bibnamefont {Coles}}, \bibinfo {author}
  {\bibfnamefont {B.}~\bibnamefont {Royall}}, \bibinfo {author} {\bibfnamefont
  {E.}~\bibnamefont {Clarke}}, \bibinfo {author} {\bibfnamefont
  {C.}~\bibnamefont {Bentham}}, \bibinfo {author} {\bibfnamefont
  {N.}~\bibnamefont {Prtljaga}}, \bibinfo {author} {\bibfnamefont {I.~E.}\
  \bibnamefont {Itskevich}}, \bibinfo {author} {\bibfnamefont {L.~R.}\
  \bibnamefont {Wilson}}, \bibinfo {author} {\bibfnamefont {M.~S.}\
  \bibnamefont {Skolnick}}, \ and\ \bibinfo {author} {\bibfnamefont {A.~M.}\
  \bibnamefont {Fox}},\ }\href {\doibase 10.1038/s41565-018-0188-x} {\bibfield
  {journal} {\bibinfo  {journal} {Nature Nanotechnology}\ }\textbf {\bibinfo
  {volume} {13}},\ \bibinfo {pages} {835} (\bibinfo {year} {2018})}\BibitemShut
  {NoStop}%
\bibitem [{\citenamefont {Tomm}\ \emph {et~al.}(2021)\citenamefont {Tomm},
  \citenamefont {Javadi}, \citenamefont {Antoniadis}, \citenamefont {Najer},
  \citenamefont {Löbl}, \citenamefont {Korsch}, \citenamefont {Schott},
  \citenamefont {Valentin}, \citenamefont {Wieck}, \citenamefont {Ludwig},\
  and\ \citenamefont {Warburton}}]{Tomm2021}%
  \BibitemOpen
  \bibfield  {author} {\bibinfo {author} {\bibfnamefont {N.}~\bibnamefont
  {Tomm}}, \bibinfo {author} {\bibfnamefont {A.}~\bibnamefont {Javadi}},
  \bibinfo {author} {\bibfnamefont {N.~O.}\ \bibnamefont {Antoniadis}},
  \bibinfo {author} {\bibfnamefont {D.}~\bibnamefont {Najer}}, \bibinfo
  {author} {\bibfnamefont {M.~C.}\ \bibnamefont {Löbl}}, \bibinfo {author}
  {\bibfnamefont {A.~R.}\ \bibnamefont {Korsch}}, \bibinfo {author}
  {\bibfnamefont {R.}~\bibnamefont {Schott}}, \bibinfo {author} {\bibfnamefont
  {S.~R.}\ \bibnamefont {Valentin}}, \bibinfo {author} {\bibfnamefont {A.~D.}\
  \bibnamefont {Wieck}}, \bibinfo {author} {\bibfnamefont {A.}~\bibnamefont
  {Ludwig}}, \ and\ \bibinfo {author} {\bibfnamefont {R.~J.}\ \bibnamefont
  {Warburton}},\ }\href {\doibase 10.1038/s41565-020-00831-x} {\bibfield
  {journal} {\bibinfo  {journal} {Nature Nanotechnology}\ }\textbf {\bibinfo
  {volume} {16}},\ \bibinfo {pages} {399} (\bibinfo {year} {2021})}\BibitemShut
  {NoStop}%
\end{thebibliography}
